\documentclass[pdflatex,iicol]{sn-jnl}
\usepackage[round]{natbib}

\usepackage{graphicx}%
\usepackage{multirow}%
\usepackage{amsmath,amssymb,amsfonts}%
\usepackage{amsthm}%
\usepackage{mathrsfs}%
\usepackage[title]{appendix}%
\usepackage{xcolor}%
\usepackage{textcomp}%
\usepackage{manyfoot}%
\usepackage{booktabs}%
\usepackage{algorithm}%
\usepackage{algorithmicx}%
\usepackage{algpseudocode}%
\usepackage{subcaption}
\usepackage{listings}%
\usepackage{csquotes}
\usepackage{hyperref}
\usepackage[colorinlistoftodos]{todonotes}
\newcommand{\Est}[1]{\hat{#1}}

\newlength{\tempdima}
\newcommand{\rowname}[1]
{\rotatebox{90}{\makebox[\tempdima][c]{\textbf{#1}}}}

\begin{document}

\title[Article Title]{An analysis of the relative effects of connectivity and coupling interactions on spin networks emulating the D-Wave 2000Q quantum annealer}


\author*[1,2]{\fnm{Jessica} \sur{Park}}\email{jlp567@york.ac.uk}

\author[1]{\fnm{Susan} \sur{Stepney}}\email{susan.stepney@york.ac.uk}
\equalcont{These authors contributed equally to this work.}

\author[2]{\fnm{Irene} \sur{D'Amico}}\email{irene.damico@york.ac.uk}
\equalcont{These authors contributed equally to this work.}

\affil*[1]{\orgdiv{Department of Computer Science}, \orgname{University of York}, \orgaddress{\city{York}, \country{UK}}}

\affil[2]{\orgdiv{Department of Physics}, \orgname{University of York}, \orgaddress{\city{York}, \country{UK}}}

\abstract{From available data, we show strong positive spatial correlations in the qubits of a D-Wave 2000Q quantum annealing chip that are connected to qubits outside their own unit cell. Then, by simulating the dynamics of three different spin networks and two different initial conditions, we then show that correlation between nodes is affected by a number of factors. The different connectivity of qubits within the network means that information transfer is not straightforward even when all the qubit-qubit couplings have equal weighting. Connected nodes behave even more dissimilarly when the couplings' strength is scaled according to the physical length of the connections (here to simulate dipole-dipole interactions). This highlights the importance of understanding the architectural features and potentially unprogrammed interactions/connections that can divert the performance of a quantum system away from the idealised model of identical qubits and couplings across the chip.}




\maketitle

\section{Introduction}\label{sec1}

Quantum computation is currently being advanced on multiple fronts, including: algorithm development, qubit realisation, device manufacturing, and error correction \citep{Ahn2002-pr, Bharti2022-iu, Harris2009-qi, Pudenz2014-oy}.
Due to the relative infancy and challenging scalability of the technology, the hardware is often difficult to control precisely, and the individual qubits can be subject to significant heterogeneity. 
Algorithms will need to be optimised based on the constraints and properties of the hardware, and this will need to be chosen, modified or built based on requirements of the software task. These processes need to be done in parallel such that the software is not being optimised based on non-optimal hardware and vice versa \citep{Bandic2022-it}.

Different physical realisations of qubits have different levels of robustness to different errors, and so different realisations may be optimal for different functions \citep{Noiri2018-ym,Osada2022-os}.
It seems likely that fabrication inhomogeneities will result in a device where different individual qubits may be optimal for different functions,
potentially allowing improved performance by careful allocation of qubits. 
Before considering how to exploit heterogeneity in the system, it is crucial to understand its sources and effects.
Here we examine how heterogeneity presents itself on a quantum chip, and how this affects the performance when running certain problems.

Section \ref{sec:dwave} gives an overview of quantum annealing and some specifics about the particular architecture that is considered in the remainder of the paper. 
Section \ref{sec:losalamos} presents an investigation in the analysis of spatial correlation that we performed on a dataset provided by Los Alamos National Laboratory \citep{Nelson2022-or}.
The results from this investigation led us to develop a spin network simulator with realistic architectures and dynamics (Section \ref{sec:connection}). 
This simulator is then used to analyse the dynamics of three spin networks each formed of 8 nodes. The time evolution dynamics are calculated for two different initial conditions and two different regimes that govern the coupling between the nodes. 
These results are discussed in Section \ref{sec:Results}.
We then close with a summary of the key conclusions and suggestions for future work to further the topic (Section \ref{sec:Conc}).

\section{Quantum Annealing and D-Wave Chimera Architecture}\label{sec:dwave}

Quantum annealing is a non-universal type of quantum computing most commonly used to find the optimal solution to a problem.
It can do this by finding the global minimum of an energy landscape that encodes the problem. Quantum fluctuation and quantum tunnelling allow the annealer to escape certain local minimal in energy landscapes.

In order to solve such optimisation problems, the cost function (to be minimised) and any associated constraints are formulated into an \textit{Ising Hamiltonian} (modelling the energy of coupled qubits). This is a quantum operator that describes the energy landscape of the system. The desired result of the annealing process is that the system reaches the ground state of this Hamiltonian, which corresponds to the optimal solution of the problem. 

The Hamiltonian that describes quantum annealing is
\begin{align}
\nonumber
H(x,s) &= \frac{A(s)}{2}\left(\sum_{i}\Est{\sigma}_{X}^{(i)}\right)+
\\ \label{eq:Anneal}
&\frac{B(s)}{2}\left(\sum_{i}h_{i}\Est{\sigma}_{Z}^{(i)}+\sum_{i>j}J_{ij}\Est{\sigma}_{Z}^{(i)}\Est{\sigma}_{Z}^{(j)}\right),
\end{align}
where $x=\{x_0, x_1, x_i...x_N\}$  is the state of the $N$-qubit system;
$s$ is normalised time;
$\Est{\sigma}_{X}^{(i)}$, $\Est{\sigma}_{Z}^{(i)}$ are the Pauli matrices acting on qubit $x_i$; 
$h_i$ and $J_{ij}$ encode the problem as qubit biases and coupling weights, and, in practice, are limited by the physical hardware graph (qubit-coupling connectivity) of the annealing device. 

Annealing occurs between physical times $t=0$ and $t=t_{f}$, normalised into an annealing fraction: $s={t}/{t_f}$, so $0 \le s \le 1$. 
$A$ and $B$ are functions of  $s$ and their relative magnitudes describe the state of the system as it moves from a general superposition state (the first term) to the solution state (the second term, the Ising Hamiltonian). 

At $t=0$ ($s=0$), the system has $A(0) \gg B(0)$:
the state starts as a general superposition of states. 
The system is  slowly annealed by increasing $B$ and decreasing $A$, until at $t=t_f$ ($s=1$) we have $A(1) \ll B(1)$. This is often referred to as freezing out the quantum fluctuations. At this point the qubits, in an ideal system, are in the ground state of the second term, that is, they are in the state representing the solution to the optimisation problem. 
The annealing process needs to happen slowly enough such that the system does finish in the ground state and not in an excited state of the Ising Hamiltonian \citep{Venegas-Andraca2018-yy}. The point at which $A(s) = B(s)$ is known as the quantum critical point (QCP), by analogy to the theory of phase transitions. 

Eqn.\ref{eq:Anneal} describes an ideal system of perfect qubits and perfect coupling.
Physical devices have limitations, imperfections and inhomogenities, however.
One major limitation of quantum annealers is qubit connectivity:
not all qubit couplings can be realised;
indeed most of the $J_{ij}$ are zero (uncoupled). Another relevant limitation is that even potential couplings can be realised only within a certain range of values and only up to a certain precision. The first restricts the coupling range, and the second is a source of unwanted noise and decoherence. Similar issues affect the qubit biases $h_i$.

 \begin{figure*}[tp]
	\centering
    \includegraphics[width=0.75\textwidth]{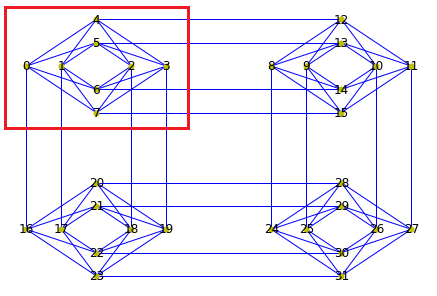}
    \caption{A graph representation of the D-Wave Chimera architecture as present on the 2000Q quantum annealer. The red box shows the 8 qubits that make up a unit cell. (Diagram created using D-Wave NetworkX Python language package \citep{D-Wave_Systems2021-bg}.)}
    \label{fig:Chimera}
\end{figure*}

Consider the D-Wave 2000Q.
It is designed with 2048 qubits in the `Chimera' architecture, which has 256 unit cells of 8 qubits, arranged in a $16 \times 16$ grid.  
There are connections between qubits inside unit cells and between qubits belonging to different unit cells. Figure \ref{fig:Chimera} shows qubit connections in a $2 \times 2$ grid of unit cells. The yellow dots in the figure represent the qubits; in reality each qubit is an elongated superconducting loop oriented either horizontally or vertically. This and the differences highlighted before may be a source of inhomogeneity in the qubit performances. The full 2000Q chip creates the $16 \times 16$ unit cells by repeating the pattern shown in Figure \ref{fig:Chimera} eight times in either dimension and connecting them in the obvious way.

How a given problem is embedded into this (and other) fixed topologies is the subject of much research.
Better characterisation of the individual qubits on the chip would allow for more intelligent and potentially real-time re-configuring embedding algorithms. 

\section{Exploring Spatial Correlations in the Los Alamos Dataset}\label{sec:losalamos}

In order to exploit maximum performance from a given quantum device, 
it is necessary to measure the performance of individual qubits and couplings in that device.
\cite{Nelson2021-ca}
perform repeated sampling of each qubit in their D-Wave 2000Q device through a range of input fields, in a process they refer to as QASA (Quantum Annealing Single-qubit Assessment).
They extract values for four parameters: inverse temperature $\beta$, bias $b$, transverse field gain $\gamma$, and noise $\eta$. 
These parameters come from a derivation in \cite{Vuffray2022-nv} that shows that states in quantum annealing platforms can be well described by a mixture of quantum Gibbs distributions that is characterised by the four parameters above. Further detail can be found in both papers cited here.

When this QASA protocol is performed for all the qubits within a chip in parallel, the variations and correlations across the chip (a 16x16 grid of unit cells) can be analysed.
The authors found that the orientation of the qubits (horizontally or vertically aligned superconducting loops) is correlated with both the inverse temperature and transverse field gain parameters. 
They hypothesise that this could be due to \enquote{asymmetry in the chip's hardware layout or to the details of how global annealing control signals are delivered to the qubits} \citep{Nelson2021-ca}.

The Los Alamos National Laboratory (LANL) research group that performed this experiment have made the raw data available,
which we use to perform further investigation into the presence of \textit{spatial correlations} in the four parameters measured for each qubit in the chip, as described in this section. 

\begin{figure*}
    \centering
    \makebox[\textwidth]{\includegraphics[width=0.7\paperwidth]{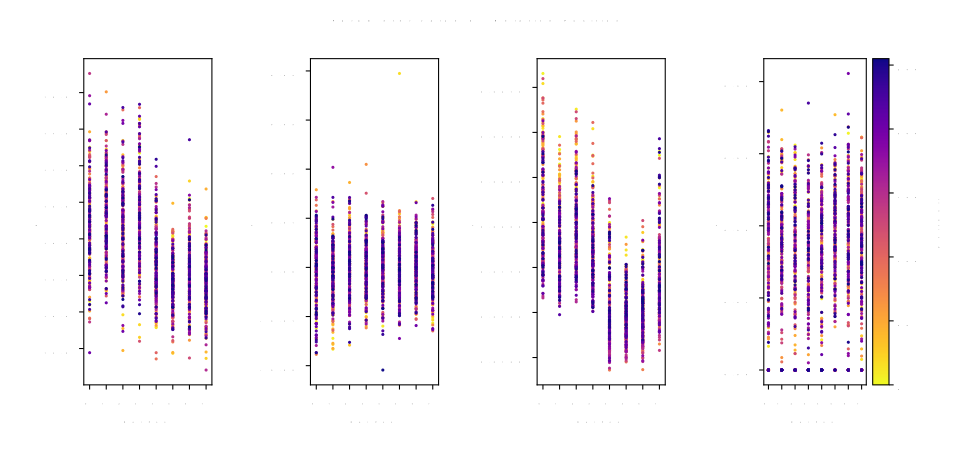}}
    \caption{Measured parameters for every qubit within a D-Wave 2000Q chip plotted by relative position in unit cell.}
    \label{fig:LANL_data}
\end{figure*}

Each qubit in a D-Wave chip can be identified either by qubit number (0-2047) or by the combination of unit cell (0-256) and relative position (0-7). For example qubit 10 labelled in Figure \ref{fig:Chimera}, could also be identified as qubit (1,2). This means it is in the second unit cell (numbering starting from 0) and within this cell, it occupies position 2 (in reference to the numbering in the first unit cell).

An initial investigation was undertaken to understand whether a qubit's relative position within its unit cell has an effect on any of the parameters extrapolated in this data set. 
The results are shown in Figure \ref{fig:LANL_data}.

It can be seen that for bias and noise, there is no significant variation between qubits at different positions in the unit cell. 
However, for inverse temperature and transverse field gain, the parameter values are significantly lower for qubit positions 4-7 compares to positions 0-3 but with little variation within these subgroups. 
These correspond to the different orientations of qubits in the chip. 
Qubits 0-3 are oriented horizontally and 4-7 are oriented vertically. 
Hence, our results agree with those presented in the original paper by Nelson \textit{et al.}. and do not find any further relationships between the qubit parameters and the position of qubits within the chip.

The colourisation of the groups in Figure \ref{fig:LANL_data} corresponds to the unit cell in which that qubit resides. Although there is significant variation within qubits in the same relative position in different unit cells, there is no significant trend between unit cell numbering and any of the parameters measured here.

We now turn to considering the connections between the qubits to investigate whether the couplings that allow neighbouring qubits to interact could be responsible for the varying parameter values seen both in the results above and in the original paper.
To measure spatial 
correlations we use Geary's $C$, a number which determines whether adjacent measurements are correlated \citep{Geary1954-cf}. By adjacent here, we mean qubits that have connections between them, either internal and external to unit cells.
$C$ is defined as:

\begin{equation}
C = \frac{(n-1)\sum_i \sum_j w_{ij}(x_i - x_j)^2}{2 \sum_i (x_i - \bar{x})^2 \sum_i \sum_j w_{ij}},
\label{eq:Gearys}
\end{equation}
where $n$ is the number of qubits,
$x_i$ is the parameter value of qubit $i$,
$\bar{x}$ is the mean value of parameter $x$,
and $w_{ij}$ is the connection weight between qubits $i$ and $j$.  We take $w_{ij} = 1$ for connected qubits,
zero otherwise \citep{Zhou2008-bj}.

$C=1$ represents no correlation, $C=0$ a perfect positive correlation, and $C>1$ an increasingly negative correlation (there is no fixed maximum values for negative correlation).  Positive correlation refers to two variables that tend to move in the same direction. For example, in this case, it would mean that a node with a low bias value tends to be connected to other nodes with low bias values. Negative correlations mean that the value of one node tends to oppose the value of its connected nodes. 

The PySAL package includes a Python script that calculates Geary's $C$, but this could not be used in this case as it requires consecutively numbered nodes \citep{Rey2010-uu}. This data has a number of `dead' qubits in the chip which are not included in the dictionaries of nodes and edges, so their indices are missing. Instead, bespoke code was written to calculate the Geary's C in the scenario with non-consecutively numbered qubits.

We calculate Geary's $C$ for the entire dataset (Table~\ref{table:QASA2},  column titled ``all").
The values are very close to 1, indicating little correlation between connected qubits in any of the parameters. This is maybe to be expected if the qubits are well isolated from one another but could also mask correlations between specific subsets of data.

We then calculated $C$ for two subsets: involving either just the connections internal to unit cells, or just between unit cells (external). 
The \enquote{all} column represents a weighted average of the \enquote{internal} and \enquote{external} columns; it was calculated using all the connections on the chip, of which there are more internal than external.

Table \ref{table:QASA2} shows that qubits that are connected \textit{between} unit cells have a strong positive correlation in the inverse temperature and transverse field gain parameters, while still rather strong but negative correlations affect internal connections within  unit cells. 
Here we label correlations as `strong' when there is more than 10\% difference from the global value found in the \enquote{all} column.

\begin{table}[tp]
    \centering
    \begin{tabular}{l@{\hspace{3mm}}c@{\hspace{3mm}}c@{\hspace{3mm}}c}
           \toprule
     		 & all & internal & external \\
            \midrule
        inverse temperature, $\beta$ 
        & 1.08 & 1.30 & 0.58  \\
        
        bias, $b$ 
         & 0.93  &  0.93 & 0.92 \\
        
        transverse field gain, $\gamma$ 
        & 1.06 & 1.40 & 0.32 \\
        
        noise, $\eta$ 
        &  0.91 &  0.91 &  0.89 \\
        \bottomrule
    \end{tabular}
    \caption{Geary's $C$ spatial auto-correlation of four parameters on the Los Alamos D-Wave 2000Q chip, for all connections, for internal only connections, and for external only connections.}
    \label{table:QASA2}
\end{table}

We might expect that internal connections would correspond to physically closer qubits, and therefore more positively correlated properties, but this does not seem to be the case for these parameters.
We do not actually know the physical distances between qubits in the D-Wave system:
the graphical representation in Figure \ref{fig:Chimera} is just a schematic,
and does not show the real lengths of the different connections. 
When more details on the physical hardware realisation become available, it will be important to confirm if physical separation distance is responsible for the observed correlation between qubits.  

\section{Effect of Connection Strengths on Dynamics}\label{sec:connection}

In the LANL QASA experiment, all the connection weights are set to zero, in order to isolate the qubits from any coupling effects. 
Nevertheless, differences are seen in correlations between internal (to the unit cell) and externally coupled qubits,
implying some holdover effect.

Here we investigate\footnote{%
Preliminary results of these investigations are reported in \citet{Park2023-zc}.
Here we extend those results to include two more networks (min-max and mid-lengths networks),
and a further initial state (superposition state).
} correlations explicitly due to coupling strengths that vary due to different coupling lengths.
Due to the planar architecture of the chip, links must be of different physical lengths in order to connect qubits both within and between unit cells. Such physical differences could contribute to differing behaviours 
of qubits. The connection weights can be scaled based on relative ratios of their representative lengths in the diagram.  
We test three small simulated networks with and without the spin-spin coupling weights having been scaled to their respective lengths.

The spins in a spin network can represent any type of qubit, including the superconducting qubits used in the D-Wave chip. A spin network is a mathematically general model for this purpose.
We have developed a spin network simulator in Python that takes as input a network (based on the Chimera qubit layout shown in fig.\ref{fig:Chimera}) and emulates the state's natural dynamics considering the network connectivity and coupling  and the initial state of the system. 

The two initial states tested in this paper look to emulate two different uses of qubit networks. 
The first emulates an information transfer application where an excitation starts localised at one node and then spreads through the network. This may be how errors could propagate through a network in a computing application. This initial state is represented by one qubit being set to $|1\rangle$ and all others to state $|0\rangle$  at $t=0$.  This will be known as the localised excitation state. As this is a closed system, we know that there will be a total of one excitation in the system at all times although the occupation probability could be spread throughout the network. 

A second initial state was tested which is an equal superposition of all states in which the excitation is localised at each site within the network. This is similar to the initial state used in quantum annealing, seen in the first term of equation \ref{eq:Anneal} and will be referred to here as the superposition state.

\subsection{Hamiltonian Generation Methodology}
The qubits in the D-Wave chip are physically implemented by rf-SQUIDs (radio frequency Superconducting Quantum-Interference Devices) and the couplings are implemented by Compound Josephson-junction rf-SQUIDs \citep{Harris2009-qi}. 
The way the physical length of a coupling affects its performance is based on the underlying physical processes. 
We chose to investigate repulsive dipole-dipole interactions, which scale with distance as 
\begin{equation}
J_{ij} \propto \frac{1}{r_{ij}^3}
\label{dip}
\end{equation}

 to represent the physical interactions taking place within the system.
Eqn. (\ref{dip}) describes well the dominant qubit-qubit interaction for various qubits' physical realisations. Other types of interaction are possible, including interactions beyond nearest neighbours, and will be subject of future investigations.
We compare against a control case where all
coupling weights are equal (corresponding to an $N$-d hypercube layout).

All the coupling strengths are scaled based on the shortest connection having a weight of~1. 
This value is chosen because when the D-Wave chip is operated under normal conditions, all the given coupling weights are rescaled to lie between $-1$ and $1$. Here, for simplicity, we analyse the case in which all couplings have the same sign.

The procedure for defining the Hamiltonian matrix of the simulation is given in Algorithm \ref{alg:HM}. The required inputs define the spin network model as a list of nodes and edges numbered according to Figure \ref{fig:Chimera}. The positions (relative coordinates) of the nodes are hard coded into the simulator based on the graphical representation of the chip shown in Figure \ref{fig:Chimera}. This section of the code produces an $N \times N$ matrix (the Hamiltonian) where the diagonal terms represent the qubit biases (the $h_i$ values in Eqn.\ref{eq:Anneal}) and the other terms are the coupling weights ($J_{ij}$). If there is an edge connecting nodes $i$ and $j$, then $0 < J_{ij} \le 1$ otherwise $J_{ij} = 0$. The resulting matrix is  symmetric: 
$J_{ij} = J_{ji}$.
In this simulation, we assume all the qubit biases to have the same value, and hence, as the total energy is defined up to a constant, they can be set to zero: $h_i = 0$. 

\begin{algorithm}[tp]
\caption{Create Hamiltonian Matrix($NodeList, EdgeList, ScalingType$)}\label{alg:HM}
\begin{algorithmic}[1]
\State $ds$ := EucLengths(EdgeList) 	\Comment{Distances; Edge lengths are Euclidean distance between the nodes}
\State NodeList, EdgeList := Remap(NodeList, EdgeList) 	\Comment{Remap from native qubit indices to ordered range (0,N)}
\State M := 2D array of size (N, N)	\Comment{Initialise the Hamiltonian matrix}
    \For{idx, item in $EdgeList$}
    
    \If{ScalingFactor = Constant}
        \State $J := J0$
    \ElsIf{ScalingFactor = Dipole}
        \State $J := J0 \cdot (min(ds)/ds[idx])^3$
     \EndIf
     
     \State M[item[0], item[1]] := J
     \State M[item[1], item[0]] := J 
     \EndFor
  \State \text{\textbf{return} M}
  \end{algorithmic}
  \end{algorithm}

\begin{figure}[tp]
\centering
\begin{subfigure}[c]{0.49\textwidth}
            \centering
            \includegraphics[width=\textwidth]{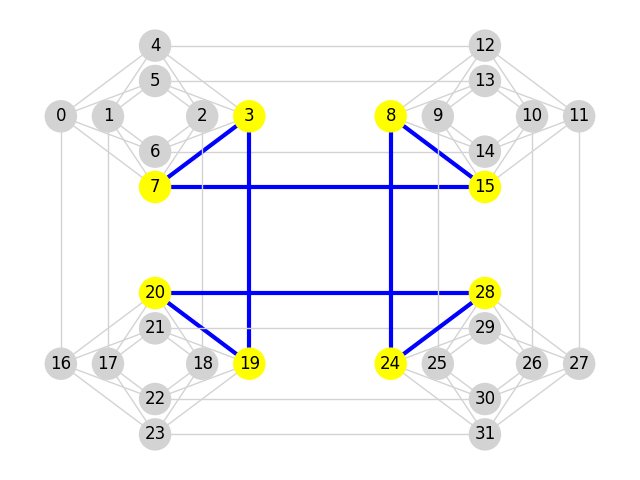}
            \caption{\small 
            Max Lengths Network\\\vspace{1em}}
\label{fig:8N_UCNC}
\end{subfigure}
        \hfill
\begin{subfigure}[c]{0.49\textwidth}
            \centering
            \includegraphics[width=\textwidth]{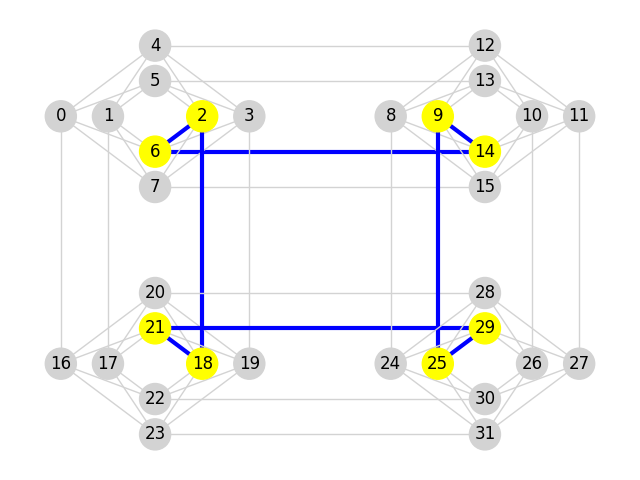}
            \caption{\small 
            Min-Max Network\\\vspace{1em}}
            \label{fig:8N_New}
            \end{subfigure}
        \hfill
        \begin{subfigure}[c]{0.49\textwidth}  
            \centering 
            \includegraphics[width=\textwidth]{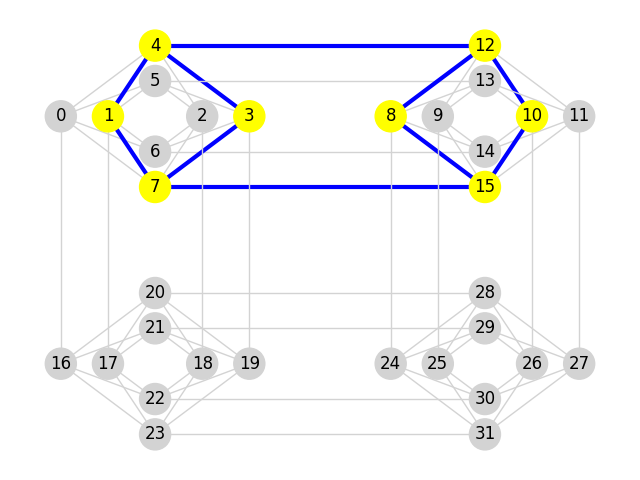}
            \caption{\small 
            Mid Lengths network}
            \label{fig:8N_Irene}
            \end{subfigure}
\caption{Test networks of Chimera topology.}
\label{fig:Models}
\end{figure}

Three networks of different structure are used for this investigation. 
All 3 networks are based on a 4 unit cell arrangement of the Chimera architecture and are shown in Figure \ref{fig:Models}. 
With three networks and two coupling scenarios for each, six Hamiltonians are produced in total. 

All the networks tested use three different edge lengths with the scaling such that the smallest is always equal to length 1. The first, referred to as \enquote{Max lengths Network}, uses the three longest lengths in the Chimera architecture, 1 internal length and the 2 external lengths, their scaled lengths are shown in Table \ref{table:MaxLengths}. The second network known as \enquote{Min-Max Network}, uses the shortest available internal connection and the 2 long external connections (scalings shown in Table \ref{table:MinMax}). The third and final network, \enquote{Mid Lengths Network} uses the lengths shown in Table \ref{table:MidLength}. These are two median lengths and the long horizontal connection.

\begin{table}[tp]
	 \centering
  \captionsetup{width=\linewidth}
  \renewcommand{\tabcolsep}{0.15cm}
	 \begin{tabular}{lcc}
	 	 \toprule
	 	 Distance & Constant & Dipole \\ 

	 	 \midrule
1.000&1.000&1.000\\
2.174&1.000&0.097\\
2.920&1.000&0.004\\

	 	 \bottomrule
	 \end{tabular}
      \hfill
	 \caption{Scaled connection lengths and calculated coupling strengths for the Max Lengths Network}
	 \label{table:MaxLengths}

	 \centering
  \renewcommand{\tabcolsep}{0.15cm}
	 \begin{tabular}{lcc}
	 	 \toprule
	 	 Distance & Constant & Dipole \\ 

	 	 \midrule
1.000&1.000&1.000\\
4.592&1.000&0.010\\
6.168&1.000&0.004\\
	 	 \bottomrule
	 \end{tabular}
      \hfill
	 \caption{Scaled connection lengths and calculated coupling strengths for the Min-Max Network}
	 \label{table:MinMax}

	 \centering
  \renewcommand{\tabcolsep}{0.15cm}
	 \begin{tabular}{lcc}
	 	 \toprule
	 	 Distance & Constant & Dipole \\ 

	 	 \midrule
1.000&1.000&1.000\\
1.392&1.000&0.371\\
4.064&1.000&0.015\\
	 	 \bottomrule
	 \end{tabular}
      \hfill
	 \caption{Scaled connection lengths and calculated coupling strengths for the Mid Lengths Network}
	 \label{table:MidLength}
\end{table}

\subsection{Time Evolution Methodology}

The Hamiltonian matrix produced in the previous section is then used to simulate the time evolution dynamics using a method described e.g. in Mortimer \textit{et al.}\citep{Mortimer2021-ex}. This involves solving Schr\"{o}dinger's equation by expanding $|\Psi(t)\rangle$, the state of the system at any time, in terms of the eigenvectors of the Hamiltonian. This is the preferred method over time step iterations as it doesn't accumulate errors due to the state at each time being calculated directly from the initial state. From the result of this algorithm, the probability of the excitation being measured at each node at each time step can be derived as $|\langle i|\Psi(t)\rangle|^2$, with $|i\rangle$ a shorthand for the state describing the excitation being localised at node $i$. This is referred to as the fidelity of measuring an excitation at a particular node at a particular time. 
In order to consider how the node coupling affects the system dynamics and therefore the spatial correlations in the system, we define a time window within which to consider the information (excitation) transfer through the network. The time window goes from $t=0$ to $t= {1}/{J_{min}}$, where $J_{min}$ refers to the smallest coupling weight in the system.
This time window was chosen because in real quantum devices, the relevant time scales over which operations can be performed is dependent on the strength of the couplings between the qubits. 
The gating time between nearby qubits can be estimated as the inverse of their coupling strength $\sim 1/J$, so $t= {1}/{J_{min}}$ corresponds roughly to the longest gating time in the system, and we can expect the excitation to have propagated through the network by that time. Also, within this time, it is  reasonable to expect that, in  hardware designed for quantum computation, decoherence effects are still extremely low and hence the probability of errors due to additional (and unwanted) interactions remains negligible. Effects of fabrication errors can be taken into consideration within the proposed model, e.g. following Ronke \textit{et. al} \citep{Ronke2011-rp}. However before incorporating these kind of effects into our model,  more information on the hardware details would be desirable to ensure that the simulation remains as useful as possible whilst still be generally applicable.
Within this time window, we consider the excitation fidelity at two specific times: The time at which the first fidelity peak in the time window occurs; and the time at which the maximum fidelity peak (excluding the initial node) occurs. At these times, the excitation fidelity of all nodes in the system can be measured and compared to infer the correlation between connected nodes. 

\section{Results}\label{sec:Results}
\subsection{Max Length Networks}
\subsubsection{Localised Excitation State}
The first set of results presented here are the time dynamics of the Max Length Network when initialised with the localised excitation state.
\begin{figure}[tp]
\centering
\begin{subfigure}[b]{\linewidth}
            \centering
            \includegraphics[width=0.91\textwidth]{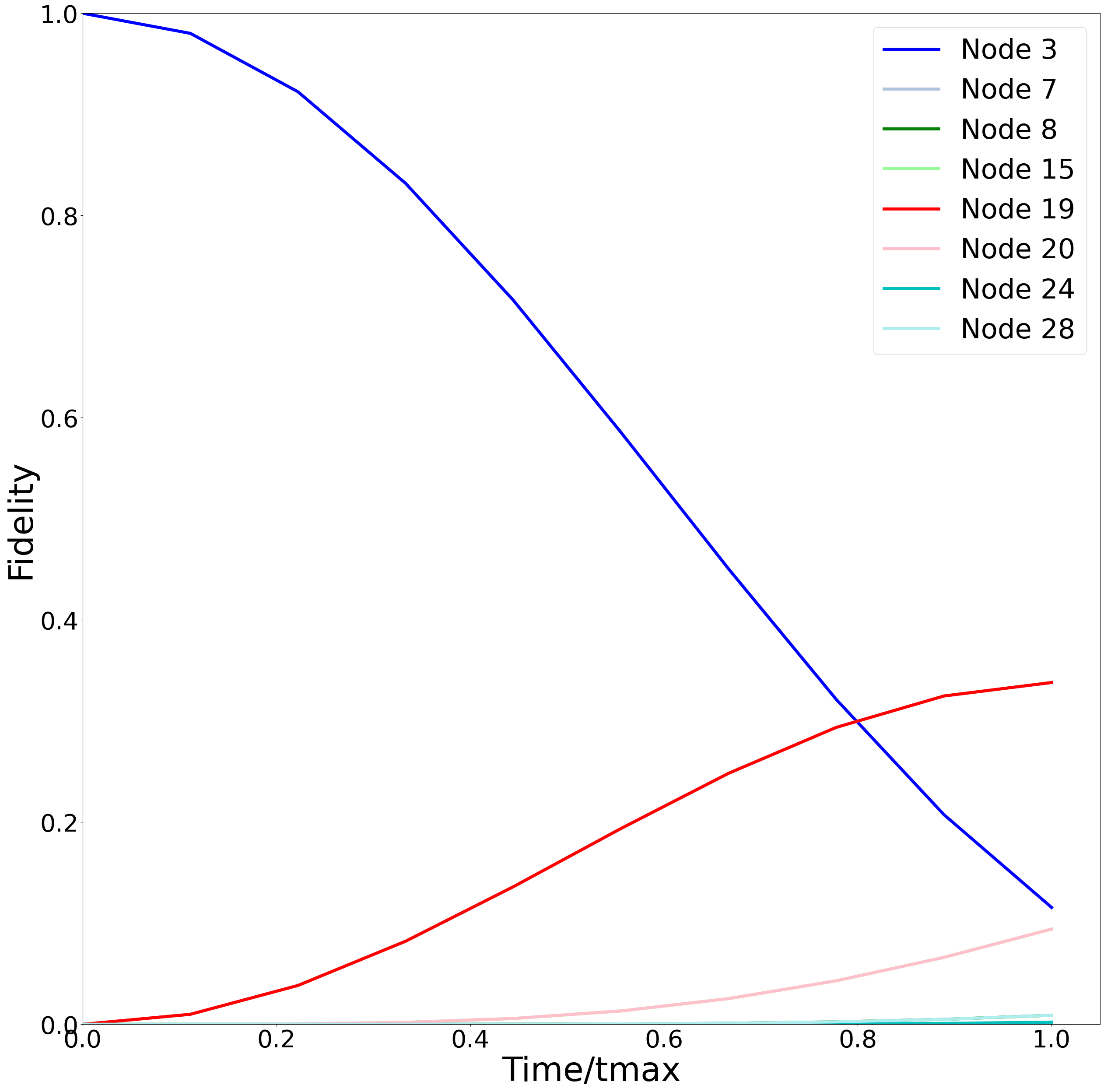}
            \caption{\small 
            coupling weights equal\\\vspace{1em}}
            \label{fig:UCNCConst}
            \end{subfigure}
        \hfill
        \begin{subfigure}[b]{\linewidth}  
            \centering 
            \includegraphics[width=0.91\textwidth]{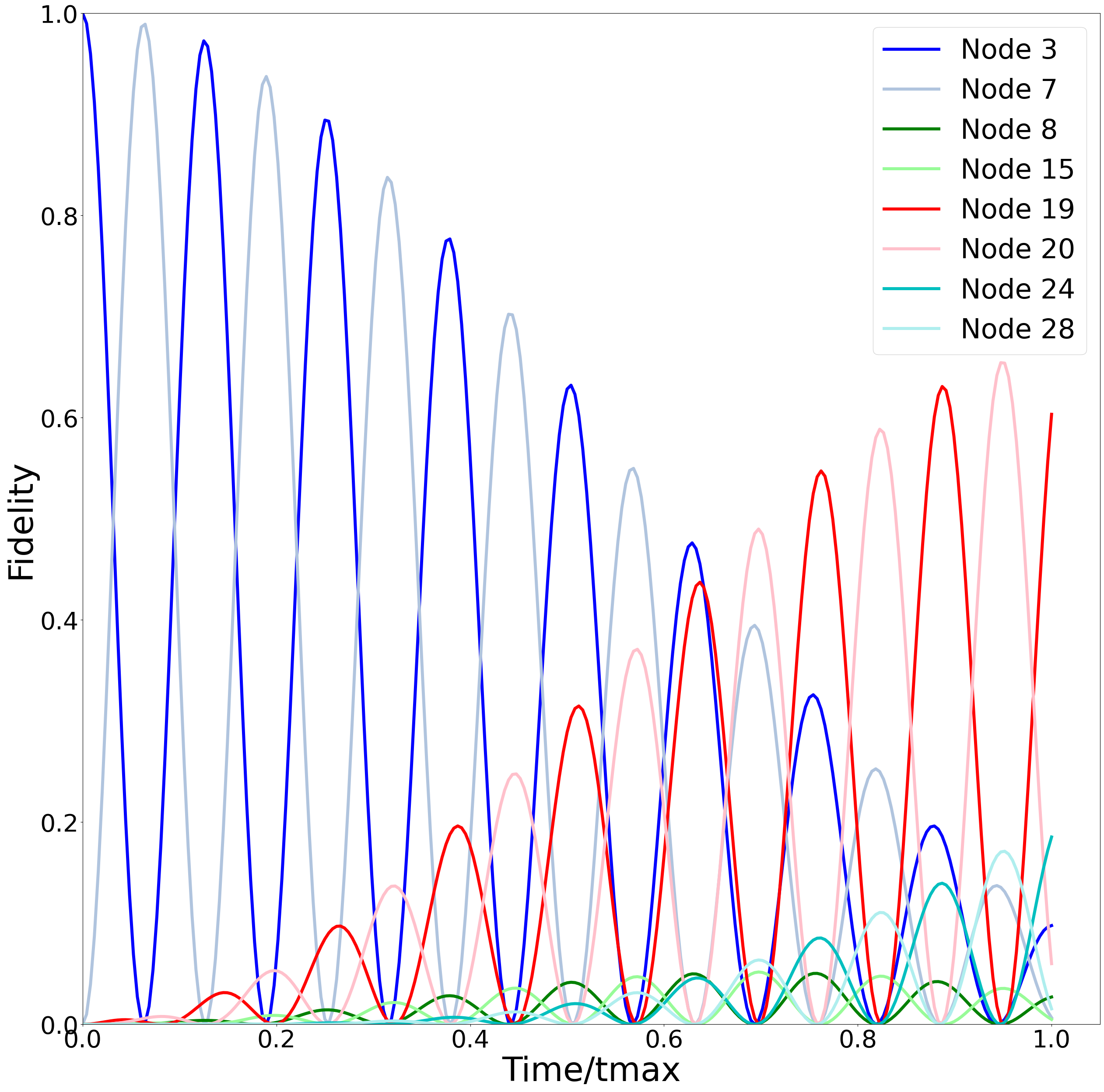}
            \caption{\small 
            coupling weights scaled like dipole-dipole interactions}
            \end{subfigure}
\caption{System dynamics of the Maximum Lengths network.}
\label{fig:8N_MaxLen}
\end{figure}

In this \enquote{Max Lengths} network (Fig \ref{fig:8N_UCNC}), each node has one internal and one external connection. All the internal connections are the same length and the external connections are either vertical or horizontal with different lengths. This means that this network is formed of the three largest lengths possible. The excitation begins on node number 3 at time $t=0$. This node is connected to nodes numbered 7 and 19 with the couplings either weighted equally, or with a dipole-dipole interaction according to their length. 

With constant (length independent) couplings we expect the fidelities of nodes \#7 and \#19 to have the same dynamics; this is shown in Figure \ref{fig:UCNCConst} with the behaviour for node \#7 being exactly overlaid by that of node \#19.

With dipole-dipole couplings, we expect nodes \#7 and \#19 to behave differently: the longer external connection here has a coupling strength of only 11\% of that of the shorter internal connection.
So the shorter connection (to node \#7) gives rise to larger fidelity peak, and  the longer (to node \#19) gives a smaller peak within the considered time-window.
This is a weak enough connection to prevent noticeable peaks in node \#19 until approximately $t = 0.2t_{max}$ allowing, initially, for a near perfect state transfer between nodes \#3 and \#7.

\begin{figure*}
\settoheight{\tempdima}{\includegraphics[width=.32\linewidth]{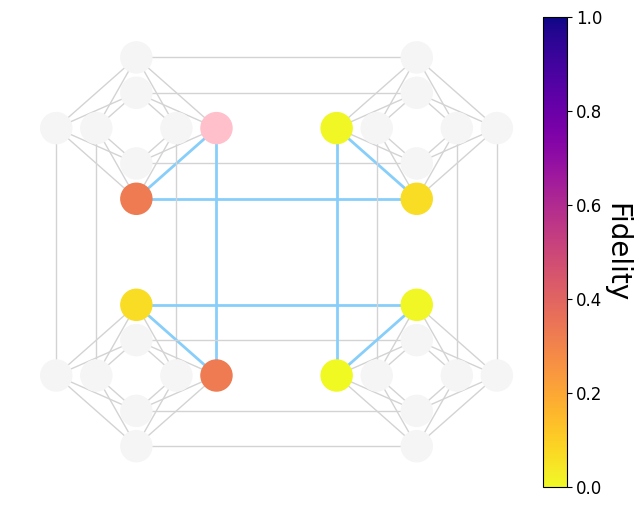}}%
\centering\begin{tabular}{@{}c@{ }c@{ }c@{}}
&\textbf{$t_{fP}$} & \textbf{$t_{mP}$}\\
\rowname{Constant}&
\includegraphics[width=.3\linewidth]{Results/UCNC/const_first.png}&
\includegraphics[width=.3\linewidth]{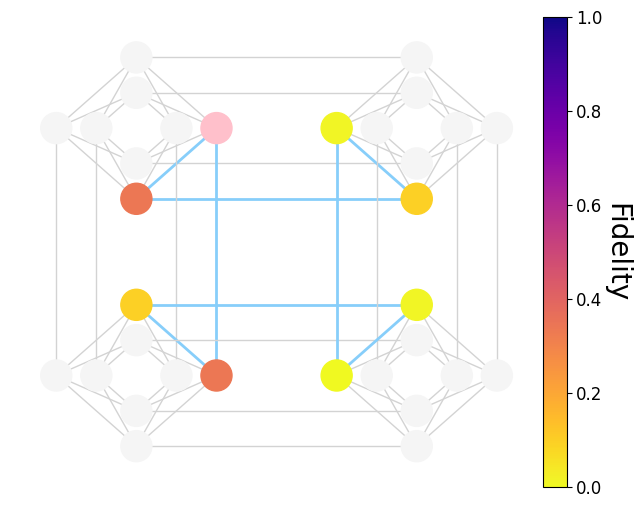}\\[-1ex]
&\textbf{(a)} & \textbf{(b)}\\
\rowname{Dipole}&
\includegraphics[width=.3\linewidth]{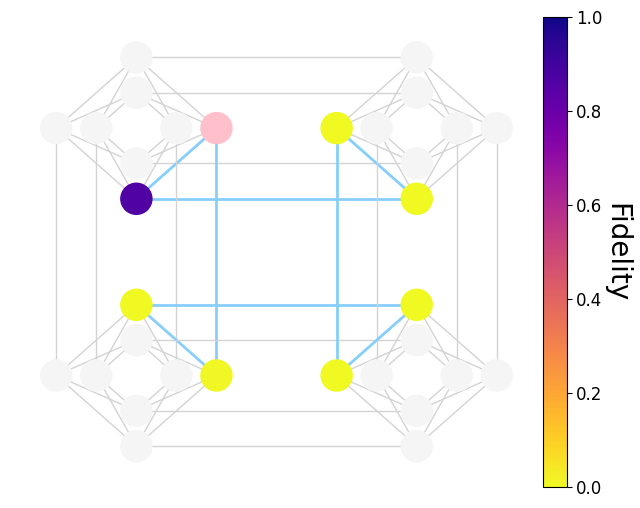}&
\includegraphics[width=.3\linewidth]{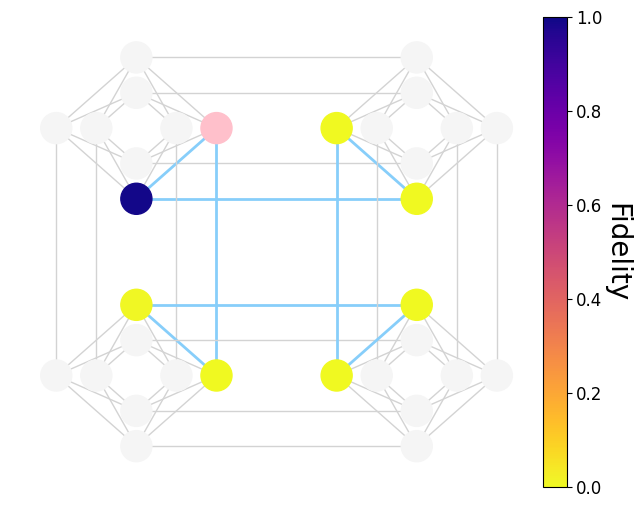}\\[-1ex]
&\textbf{(c)} & \textbf{(d)}\\
\end{tabular}
\caption{\small Node fidelities at two different times with two different couplings.}%
\label{fig:MaxLenGraphs}
\end{figure*}

To investigate  potential spatial correlations, we show the results for these network, at $t=maxPeak$ and $t=firstPeak$ in Figure \ref{fig:MaxLenGraphs}. The pink node indicates the location of the initial excitation. In the first peak dynamics, it is clear that when the nodes have constant coupling, the edges (3,7) and (3,19) behave identically. When there are dipole-dipole interactions, there is large difference in excitation transfer across these connections with the short connection producing the highest excitation transfer within the observed time window. 

\begin{figure*}
\settoheight{\tempdima}{\includegraphics[width=.32\linewidth]{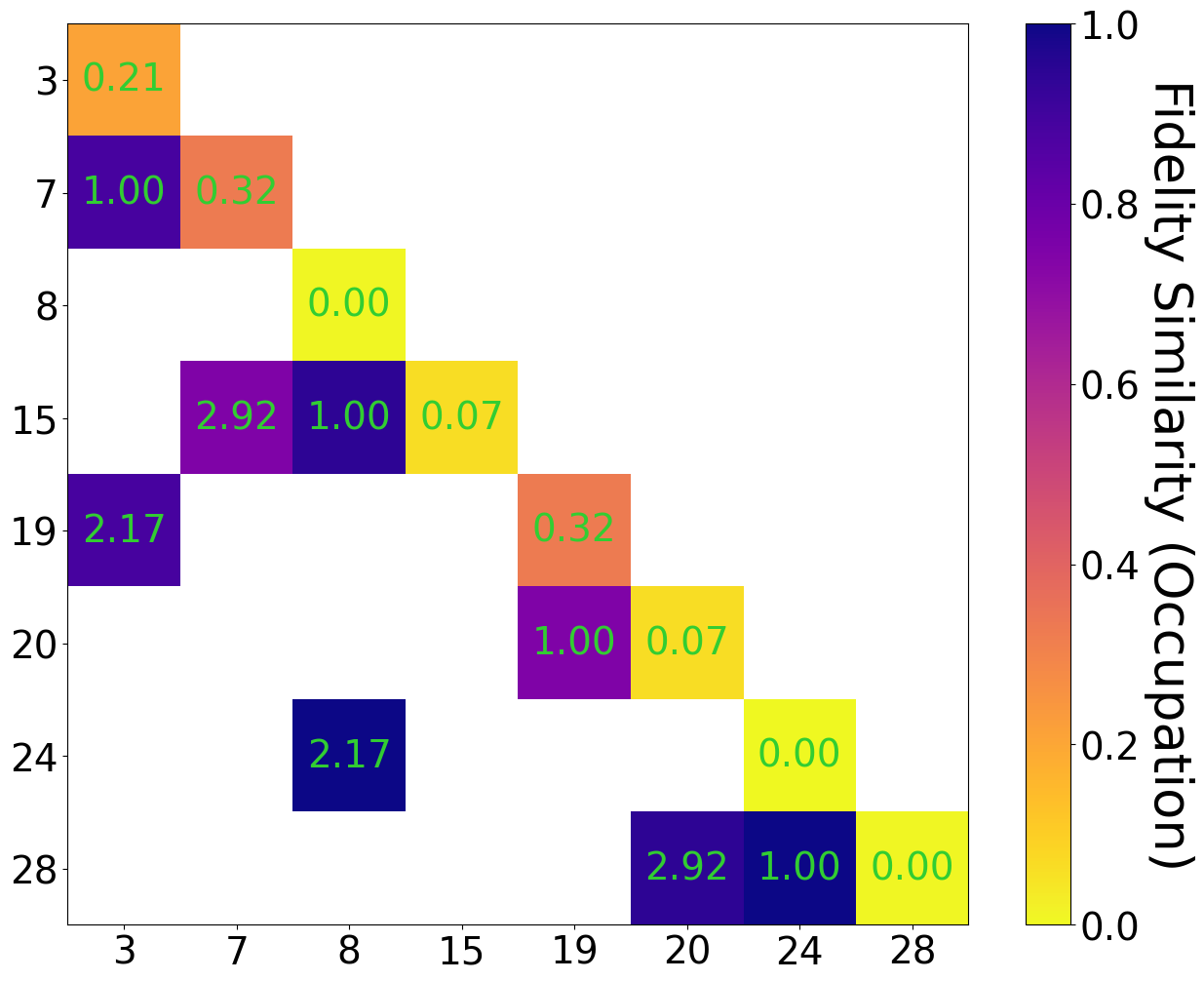}}%
\centering\begin{tabular}{@{}c@{ }c@{ }c@{}}
&\textbf{$t_{fP}$} & \textbf{$t_{mP}$}\\
\rowname{Constant}&
\includegraphics[width=.3\linewidth]{Results/UCNC/const_first_grid.png}&
\includegraphics[width=.3\linewidth]{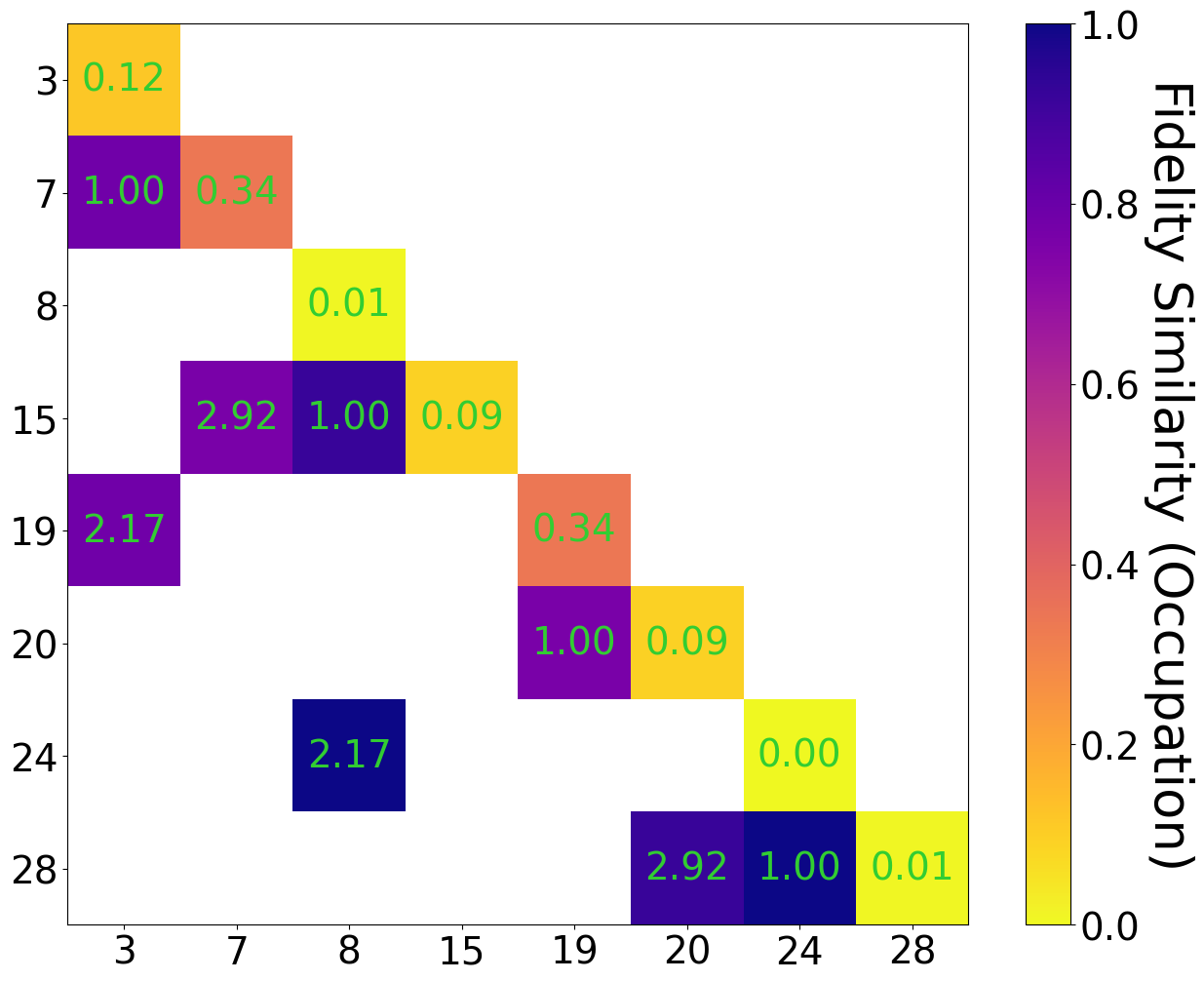}\\[-1ex]
&\textbf{(a)} & \textbf{(b)}\\
\rowname{Dipole}&
\includegraphics[width=.3\linewidth]{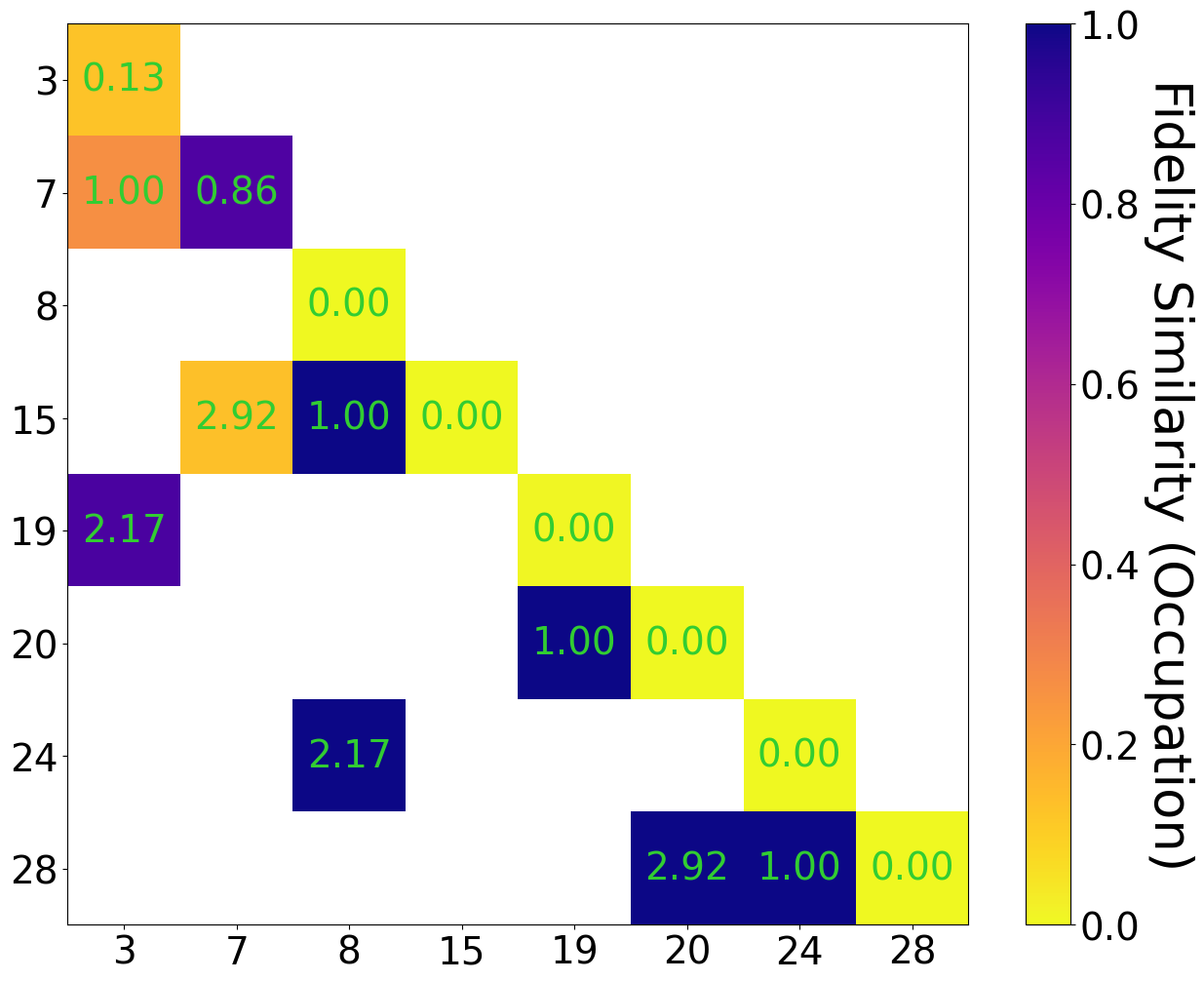}&
\includegraphics[width=.3\linewidth]{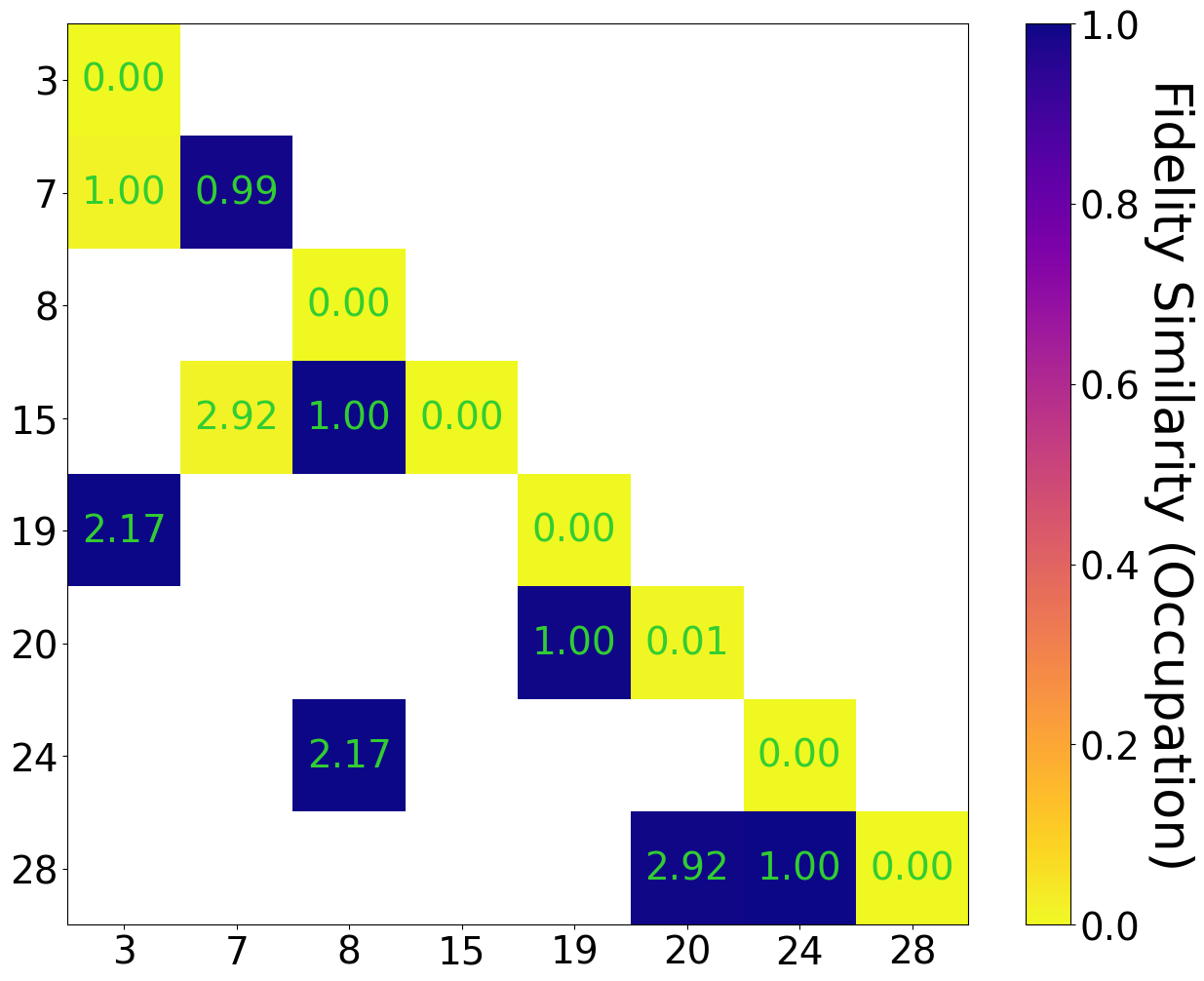}\\[-1ex]
&\textbf{(c)} & \textbf{(d)}\\
\end{tabular}
\caption{\small The similarity in the fidelity of connected nodes in the Max Lengths Network. A square at ($i,j$) is labelled with the relative length of the connection between the nodes $i$ and $j$ and is coloured by the similarity as defined by Equation \ref{sim}. The diagonal is both labelled and colourised by that node's occupation fidelity at the time in question.}%
\label{fig:MaxLenGrid}
\end{figure*}

As well as considering the overall dynamics, to better compare this simulation to the LANL experiments, we also compared the fidelities of all connected nodes. These results are shown in Figure \ref{fig:MaxLenGrid}, where each square represents the edge connecting the nodes labelled at its $x$ and $y$ positions. These squares are then coloured by the similarity in the fidelities of the nodes at either end of this edge and are labelled with the normalised connection lengths for reference. 
The similarity is defined here as, 
\begin{equation}
sim=1-|f_{i} - f_{j}|
\label{sim}
\end{equation}
where $f_{i}$ and $f_{j}$ represent the fidelities of the i$^{th}$ and j$^{th}$ node respectively. Therefore if two nodes have similar fidelities, the similarity value is maximum.

At $t=firstPeak$ it is only relevant to consider the top left of the charts as the fidelity of all but the closest 5 nodes from node 3 are all still very close to 0 which means that the similarity between connected nodes is very close to 1. 

In the constant case, we expect the length of the connection to have no effect on the similarity between the connect nodes. Although this is the case in the first column of Figure \ref{fig:MaxLenGrid}, in the other columns, this is not the case. We suggest that this is an effect of the fidelity being comprised of the occupation probability corresponding to all the different paths that the excitation could have taken from one node to the other. This is a direct effect of the connectivity of the network. 

In the dipole-dipole case, the results are further complicated by the changing connection strengths between the nodes. Intuitively we would expect that a shorter connection length would cause a higher degree of similarity between the nodes. This is not seen at either of the time steps chosen for evaluation here. The charts from the dipole-dipole simulation are noticeably different from the constant case showing that couplings that are affected by physical length will affect the spatial correlations in the system. Because the connectivity is the same in both simulations, the differences must be due to the couplings.

The difference between the constant and dipole-dipole couplings at $t=maxPeak$ is that in the dipole-dipole case, the excitation fidelity is much more concentrated at a small number of nodes meaning that the similarity of these nodes with the others is particularly low. In the constant coupling case, the excitation fidelity is more evenly spread (as seen in Figure \ref{fig:MaxLenGraphs}) which means that neighbouring nodes have higher spatial correlation. 

\subsubsection{Equal Superposition Initial State}

\begin{figure}[tp]
\centering
\begin{subfigure}[b]{\linewidth}
            \centering
            \includegraphics[width=0.91\textwidth]{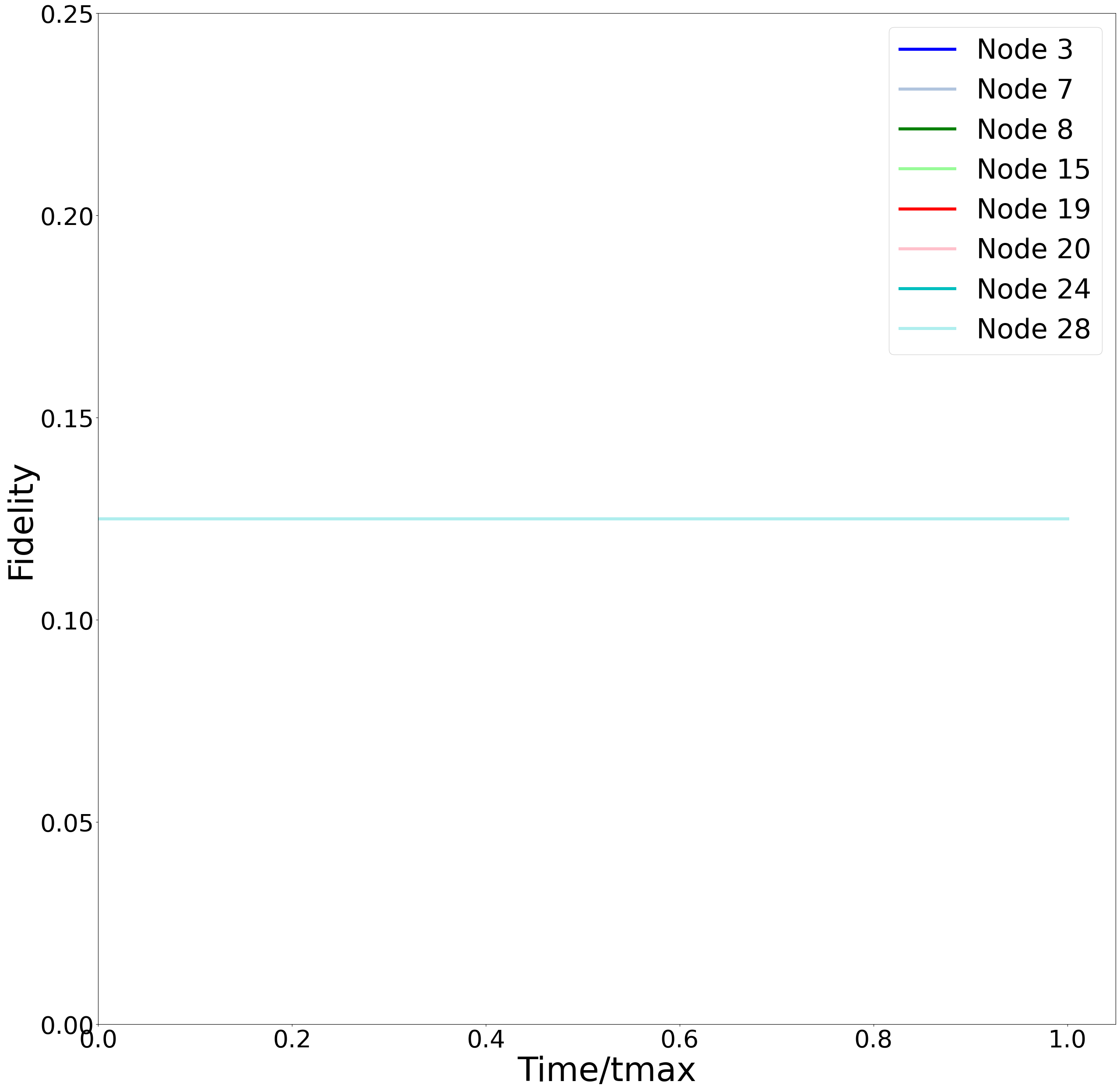}
            \caption{\small 
            coupling weights equal\\\vspace{1em}}
            \label{fig:UCNC_Equ_Const}
            \end{subfigure}
        \\
        \begin{subfigure}[b]{\linewidth}  
            \centering 
            \includegraphics[width=0.91\textwidth]{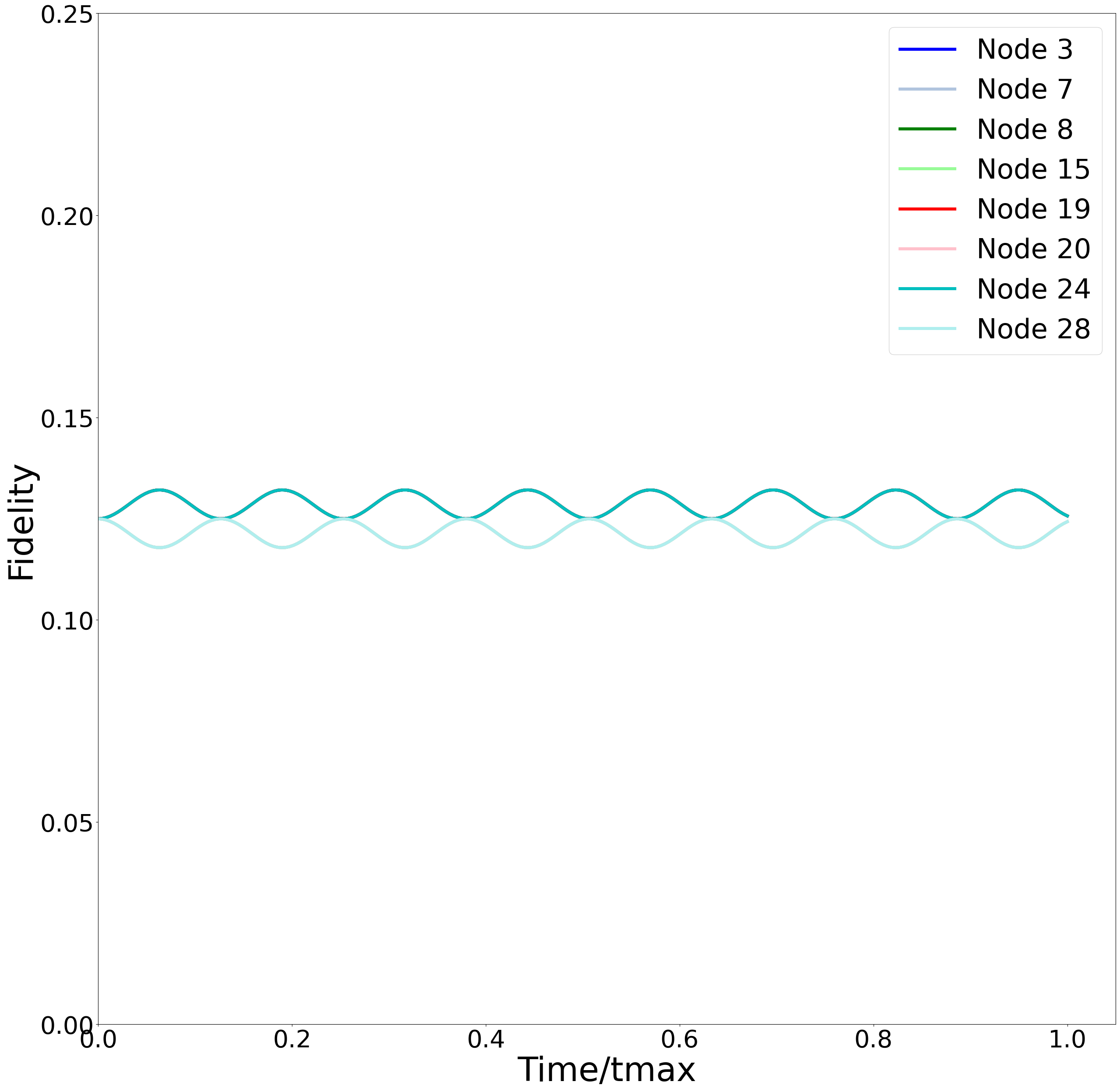}
            \caption{\small 
            coupling weights scaled like dipole-dipole interactions}
            \end{subfigure}
\caption{System dynamics of the Maximum Lengths network when initialised with the computational state.}
\label{fig:8N_MaxLen_CompState}
\end{figure}

When the Max Lengths networks is initialised with the equal superposition of single excitation states the resulting dynamics (Figure \ref{fig:8N_MaxLen_CompState}) show that in the constant coupling scenario, the state of the system is static. This is because the equal superposition state is an eigenstate of the Hamiltonian.
This network has a high degree of symmetry (when all the nodes and connections are considered equal) and only a single loops so it is not surprising that there are no dynamics within the system. 

In the dipole-dipole scenario, there are some oscillatory dynamics between pairs of qubits connected via a short connection. This is likely due to the different coupling strengths having broken one of the symmetries in the network. 

\subsection{Min-Max Network}
\subsubsection{Localised Excitation State}

\begin{figure}[tp]
\centering
\begin{subfigure}[b]{\linewidth}
            \centering
            \includegraphics[width=0.91\textwidth]{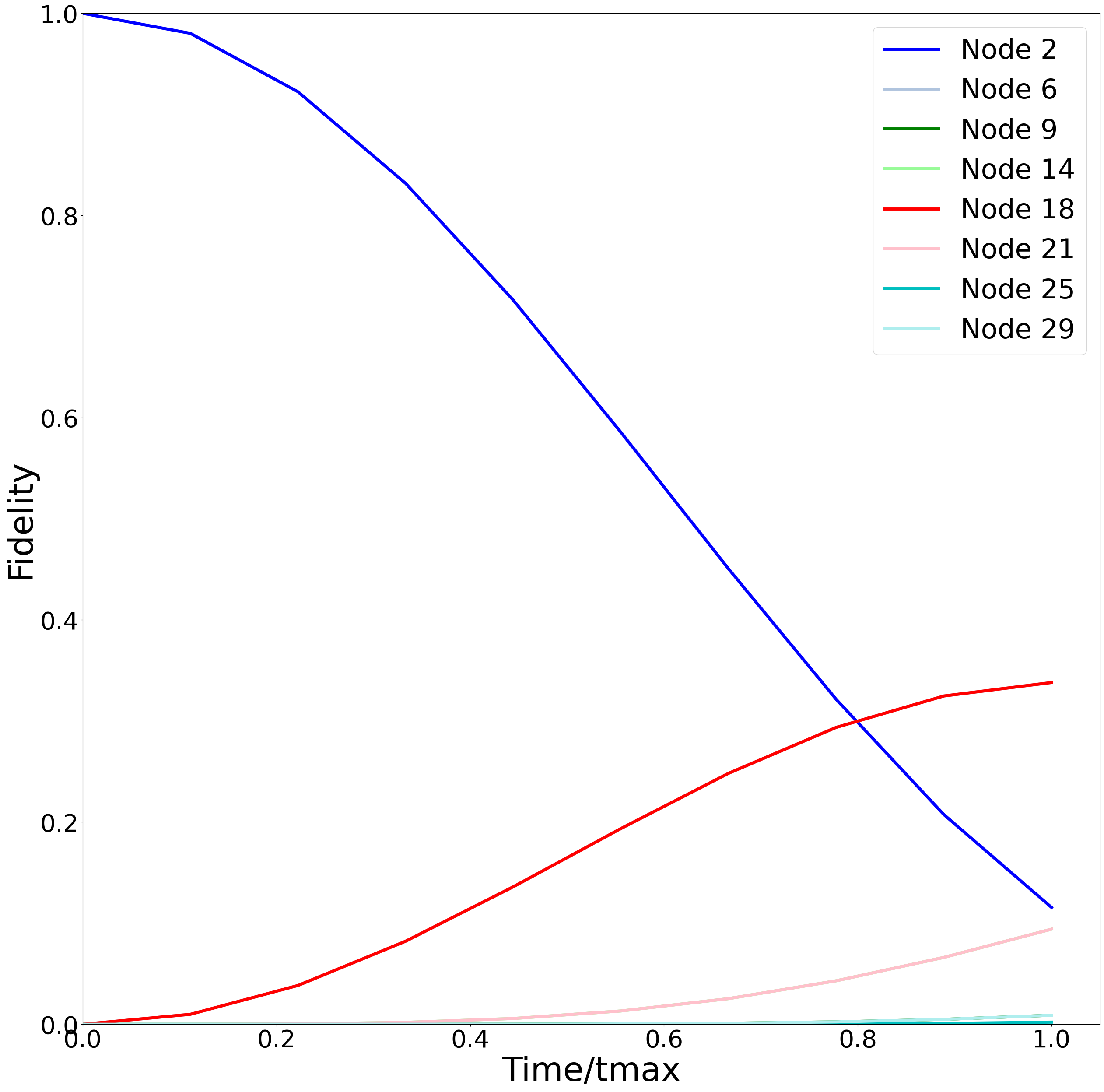}
            \caption{\small 
            coupling weights equal\\\vspace{1em}}
    \label{fig:8N_MinMaxConstant}
            \end{subfigure}
        \\
        \begin{subfigure}[b]{\linewidth}  
            \centering 
            \includegraphics[width=0.91\textwidth]{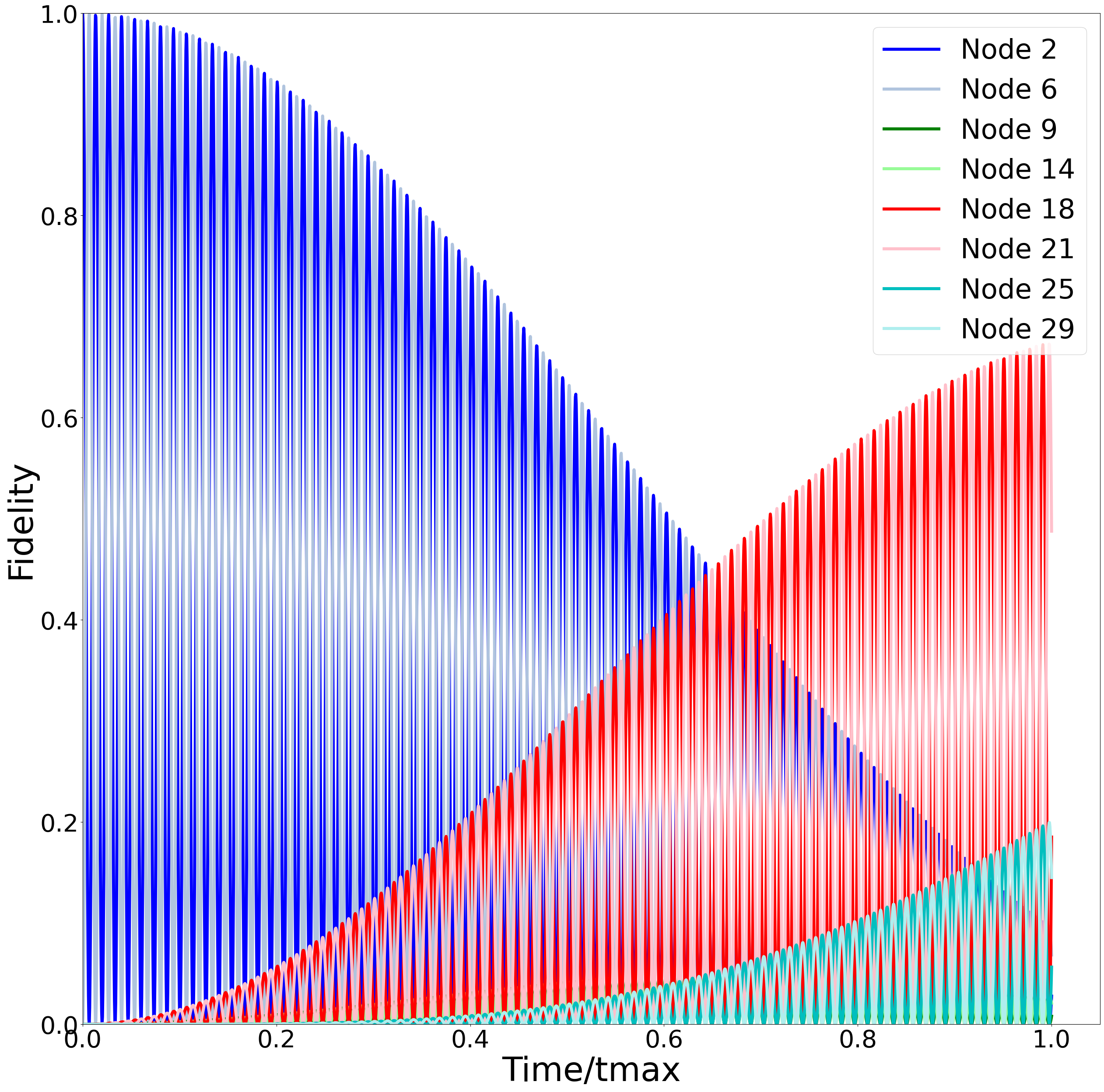}
            \caption{\small 
            coupling weights scaled like dipole-dipole interactions}
            \label{fig:MinMaxDipole}
            \end{subfigure}
\caption{System dynamics of the Min-Max network.}
\label{fig:8N_MinMax}
\end{figure}

In the Min-Max network (Fig.\ref{fig:8N_New}), each node has one internal and one external connection. All the internal connections have the same (shortest possible) length and the external connections are either vertical or horizontal, having the two longest lengths (see Table \ref{table:MinMax}). 
The simulator described in the previous section is now used to produce time dynamics in which the excitation begins localised on node number 2 at time $t=0$. This node is then connected to nodes \#6 and \#18 with the couplings either weighted equally, or with a dipole-dipole interaction according to their length. 
The time dynamics for the two scenarios are shown in Figure~\ref{fig:8N_MinMax}. 

In Figure \ref{fig:8N_MinMaxConstant}, the constant coupling means that the lines for nodes \#6 and \#18 are exactly overlaid, as are the lines corresponding to nodes \#14 and \#21, and those for nodes \#9 and \#25. These pairs represent nodes that are the same numbers of \enquote{hops} away from the site of the initial excitation. Aside from the numbering of the nodes, these dynamics are identical to those in the constant coupling case of the Max Lengths network (see Figure \ref{fig:8N_MinMax}). This is because the Hamiltonian of any single loop network has the same structure (when the nodes are identical) and the only difference between these networks is the lengths of the edges which in the constant coupling case aren't taken into account. 

With dipole-dipole couplings (Figure \ref{fig:MinMaxDipole}), we expect nodes \#6 and \#18 to behave differently: the longer external connection here has a coupling strength of only 1\% of that of the shorter internal connection. Indeed, at the beginning nodes \#2 and \#6 behave as isolated, with the excitation rapidly oscillating between the two (dark and light blue curves). As the excitation partly transfers to node \#18, its strong coupling to node \#21 makes it oscillate between these nodes (red and pink curves) with the same frequency (related to the inverse of the coupling) as the oscillations between nodes \#2 and \#6. This repeats between nodes \#25 and \#29 when the excitation reaches node \#29 (cyan and light cyan curves). Overall, the shorter connection (to node \#6) gives rise to larger fidelity peaks, and  the longer (to node \#18) gives smaller peaks within the considered time-window.

The two networks discussed thus far have a high degree of symmetry and a cyclic nature which means any excitation transfer to a node could have come via a number of different routes. They are both topologically equivalent to a single loop. The excitation could travel around the network both clockwise and anticlockwise passing through each connection once, as well as in any combination of \enquote{backwards} and \enquote{forwards} steps. 
More exactly, since this is a quantum system, the fidelities correspond to the probability of the excitation being measured at the node in question at each time step. The fidelity for each node at each time step includes all the possible routes that the excitation could have taken to be measured at that node.

\begin{figure*}
\settoheight{\tempdima}{\includegraphics[width=.32\linewidth]{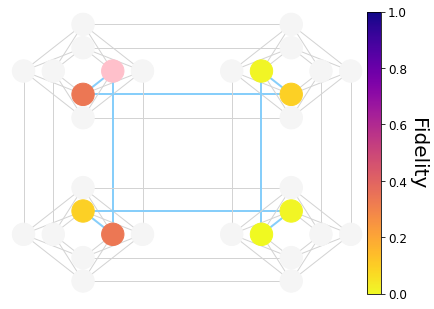}}%
\centering\begin{tabular}{@{}c@{ }c@{ }c@{}}
&\textbf{$t_{fP}$} & \textbf{$t_{mP}$}\\
\rowname{Constant}&
\includegraphics[width=.3\linewidth]{Results/MinMax/const_first.png}&
\includegraphics[width=.3\linewidth]{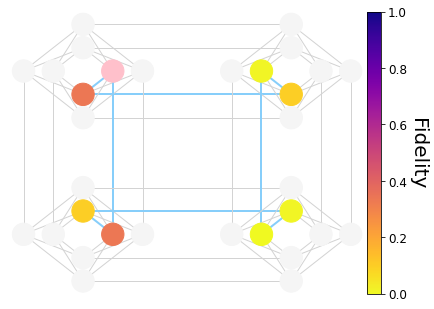}\\[-1ex]
&\textbf{(a)} & \textbf{(b)}\\
\rowname{Dipole}&
\includegraphics[width=.3\linewidth]{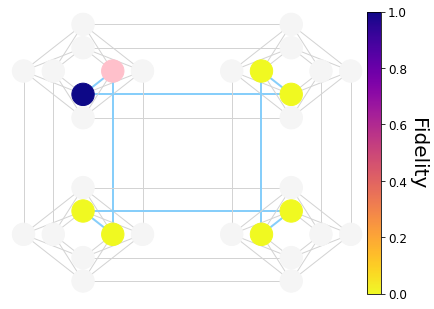}&
\includegraphics[width=.3\linewidth]{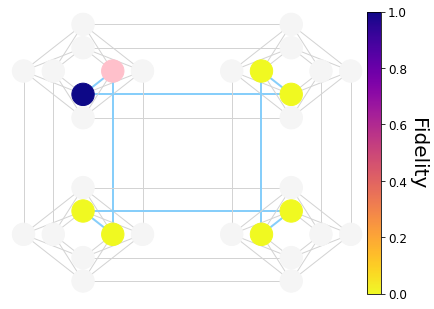}\\[-1ex]
&\textbf{(c)} & \textbf{(d)}\\
\end{tabular}
\caption{\small Node fidelities at two different times with two different couplings for the MinMax network.}%
\label{fig:MinMaxGraphs}
\end{figure*}

To investigate  potential spatial correlations, we show the results for the MinMax network, at $t=maxPeak$ and $t=firstPeak$ in Figure \ref{fig:MinMaxGraphs}. The pink node indicates the location of the initial excitation. In the first peak dynamics, it is clear that when the nodes have constant coupling, the edges (2,6) and (2,18) behave identically. When there are dipole-dipole interactions, there is large difference in excitation transfer across these connections with the short connection producing the highest excitation transfer within the observed time window. 

In the constant coupling case, the first and max peaks happen very close together in time at nodes \#6 and \#18, the immediate connections to the initial excitation reinforcing that these two edges are behaving identically. In the dipole-dipole case, we also have similar behaviours between the first and maximum peaks in that they happen very close together in time. This is to be expected in any network in which one of the connections between the injection node and adjacent noes in much bigger than the others. Compared to the constant case, these peaks are now seen only at node \#6 and the fidelity is much more concentrated.

In Figure \ref{fig:MinMaxGraphs}, subfigures a and b (constant cases), there is some fidelity at the nodes one step removed from the initial graph but this isn't seen in the dipole cases because any node one step removed has one long and one short connection and the excitation is primarily transferred by short connections.
Subfigure c (dipole-dipole coupling at the time of first peak) has more extreme peak than the equivalent in the Max Lengths graph due to the bigger difference between the short and long connections

\begin{figure*}[tp]
\settoheight{\tempdima}{\includegraphics[width=.32\linewidth]{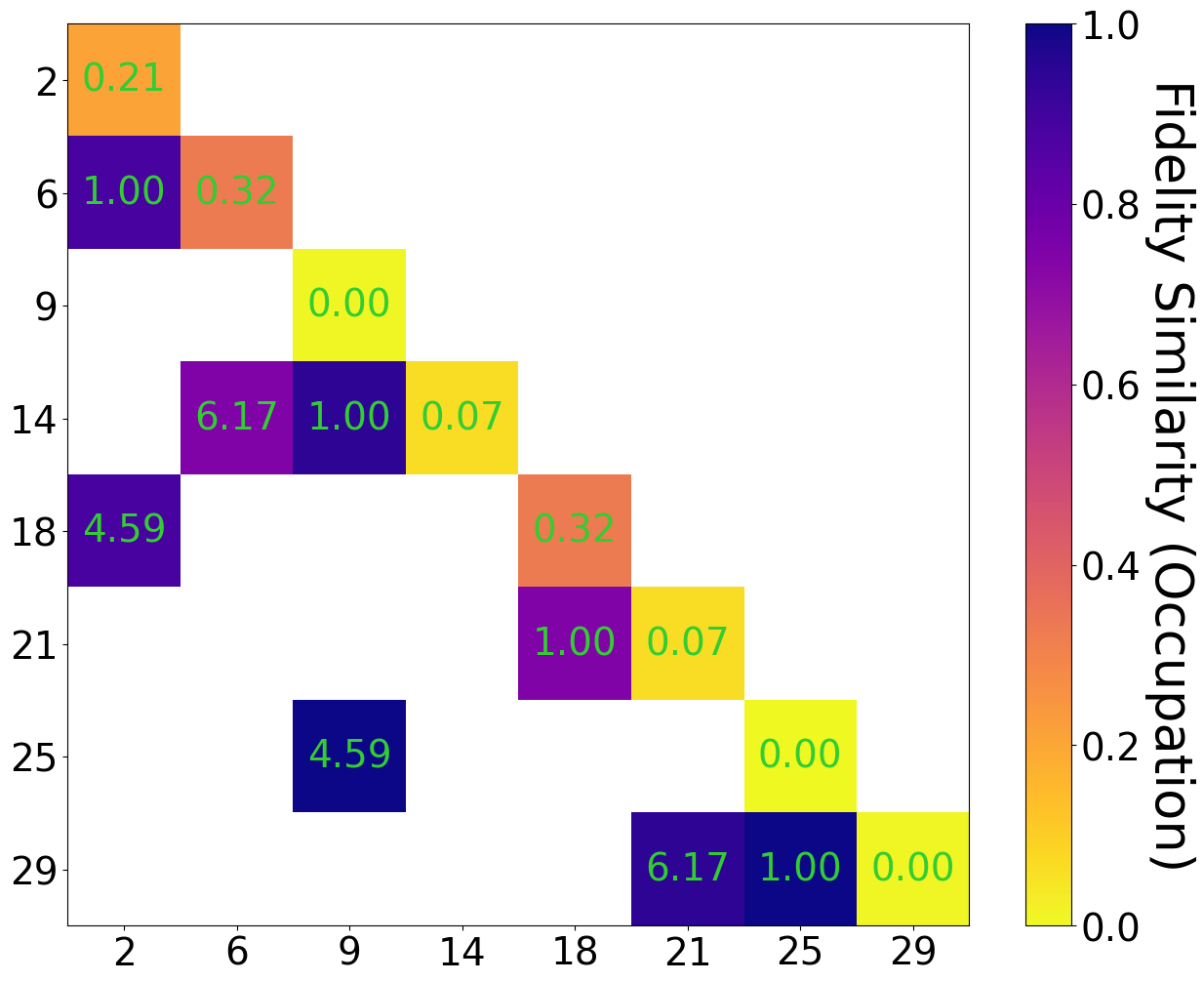}}%
\centering\begin{tabular}{@{}c@{ }c@{ }c@{}}
&\textbf{$t_{fP}$} & \textbf{$t_{mP}$}\\
\rowname{Constant}&
\includegraphics[width=.3\linewidth]{Results/MinMax/const_first_grid.png}&
\includegraphics[width=.3\linewidth]{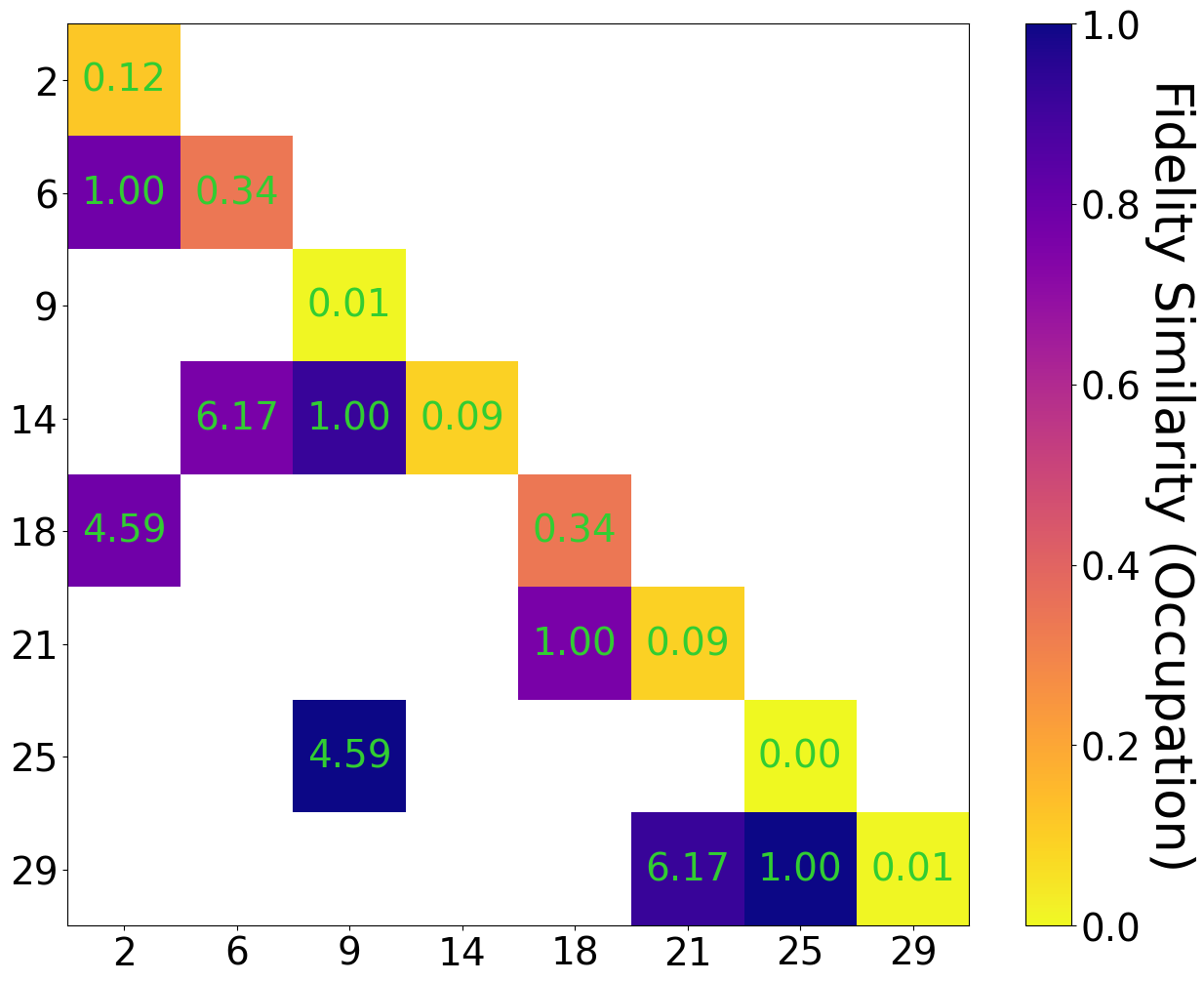}\\[-1ex]
&\textbf{(a)} & \textbf{(b)}\\
\rowname{Dipole}&
\includegraphics[width=.3\linewidth]{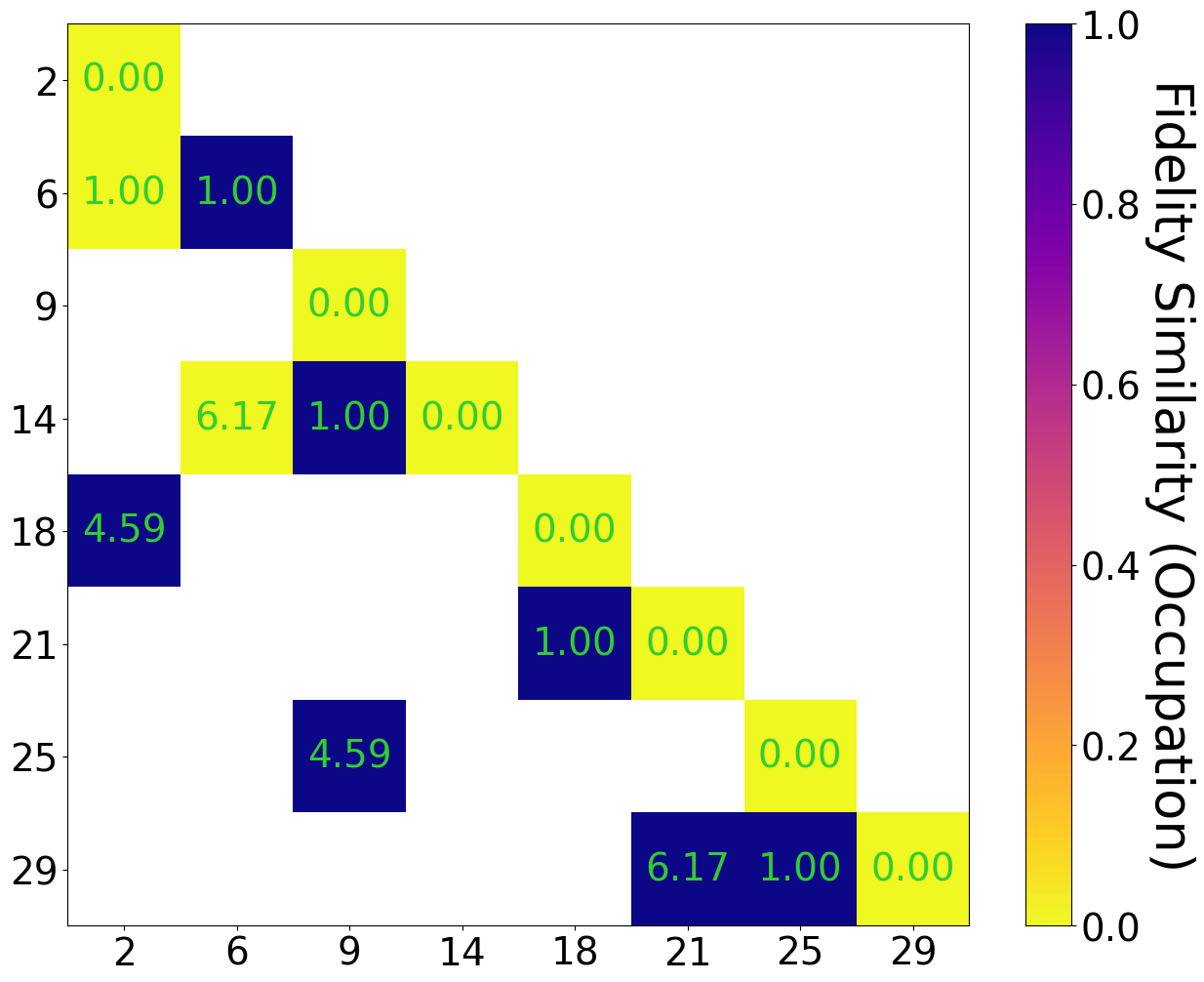}&
\includegraphics[width=.3\linewidth]{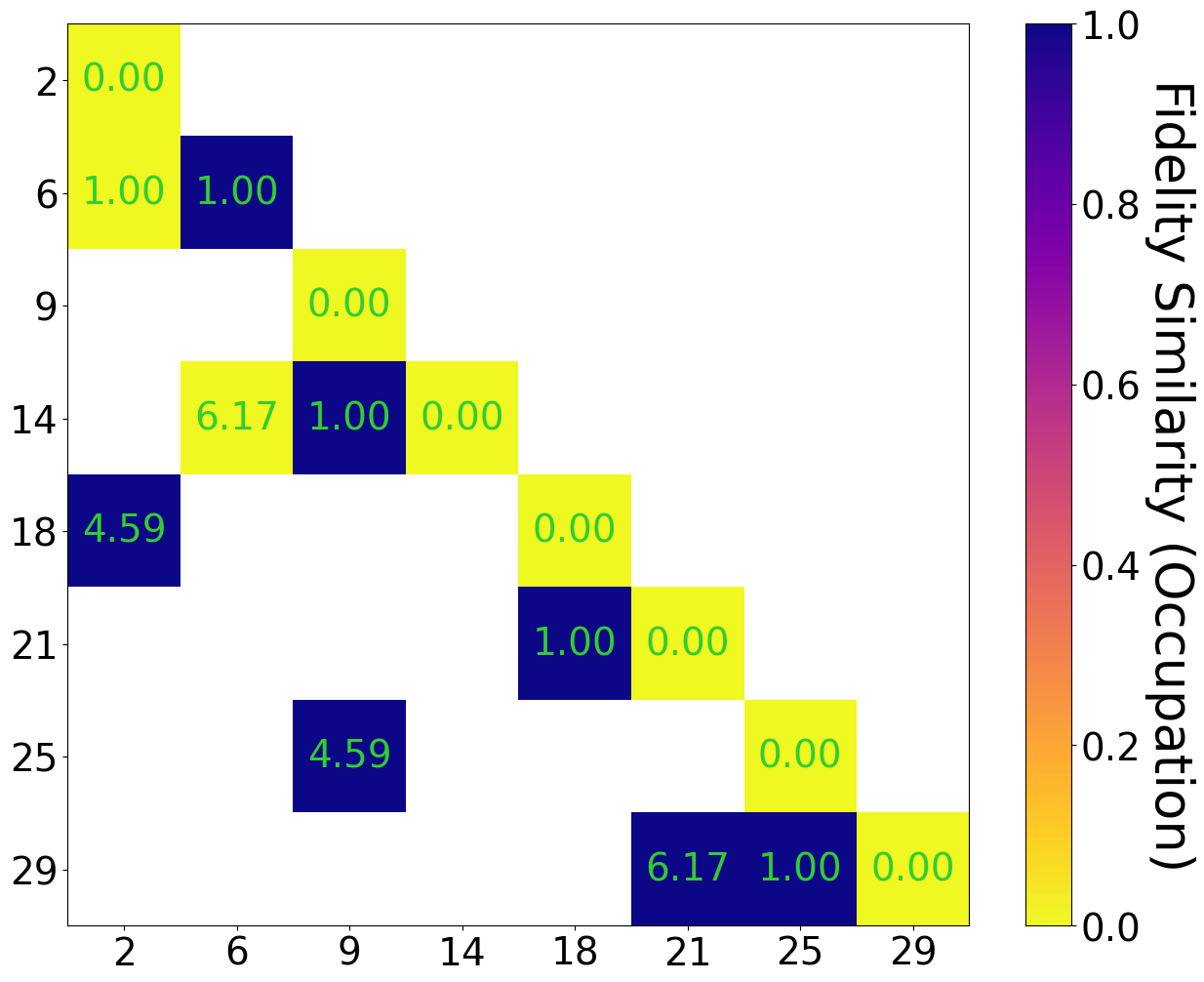}\\[-1ex]
&\textbf{(c)} & \textbf{(d)}\\
\end{tabular}
    \caption{\small  The similarity in the fidelity of connected nodes in the Min-Max Architecture. A square at ($i,j$) is labelled with the relative length of the connection between the nodes $i$ and $j$ and is coloured by the similarity as defined by Equation \ref{sim}. The diagonal is both labelled and colourised by that node's occupation fidelity at the time in question.}
    \label{fig:MinMaxGrid}
\end{figure*}

As with the previous network, we also compute the similarity between pairs of nodes. This is shown in Figure \ref{fig:MinMaxGrid} and again, it is only relevant to consider the top left of the charts as the fidelity of all but the closest five nodes from node number 2 are all still very close to zero which means that the similarity between connected nodes is very close to 1. 
In the dipole cases, the occupation probability is heavily concentrated on a single node that all the high similarity connections join nodes that have fidelities very close to zero.
In the constant case there is much more of a spread of similarity values but it is not the case that nodes are always equally similar to those they are connected to. 

\subsubsection{Equal Superposition Initial State}
\begin{figure}[tp]
\centering
\begin{subfigure}[b]{\linewidth}
            \centering
            \includegraphics[width=0.91\textwidth]{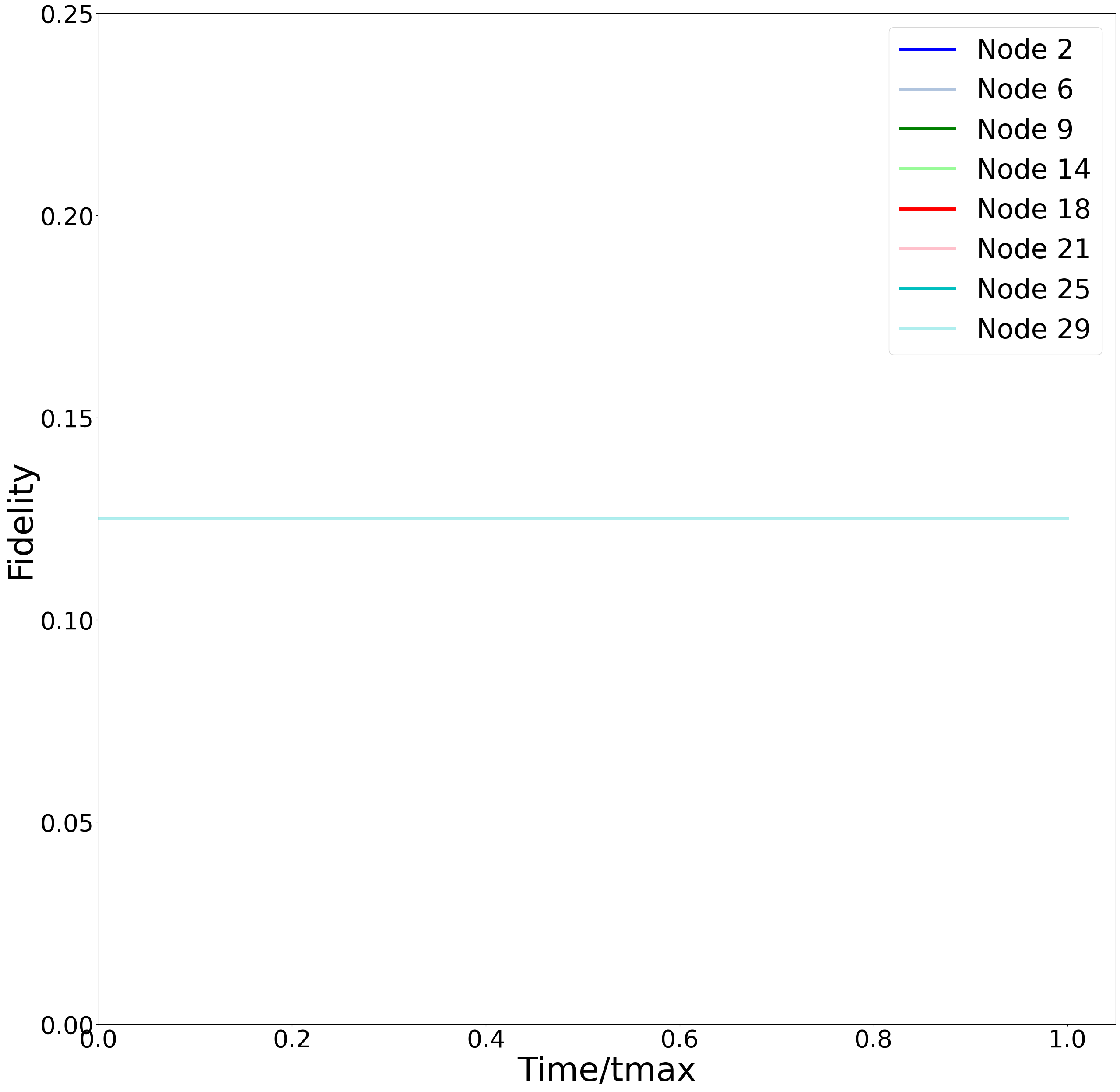}
            \caption{\small 
            coupling weights equal\\\vspace{1em}}
            \label{fig:MinMax_Equ_Const}
            \end{subfigure}
        \\
        \begin{subfigure}[b]{\linewidth}  
            \centering 
            \includegraphics[width=0.91\textwidth]{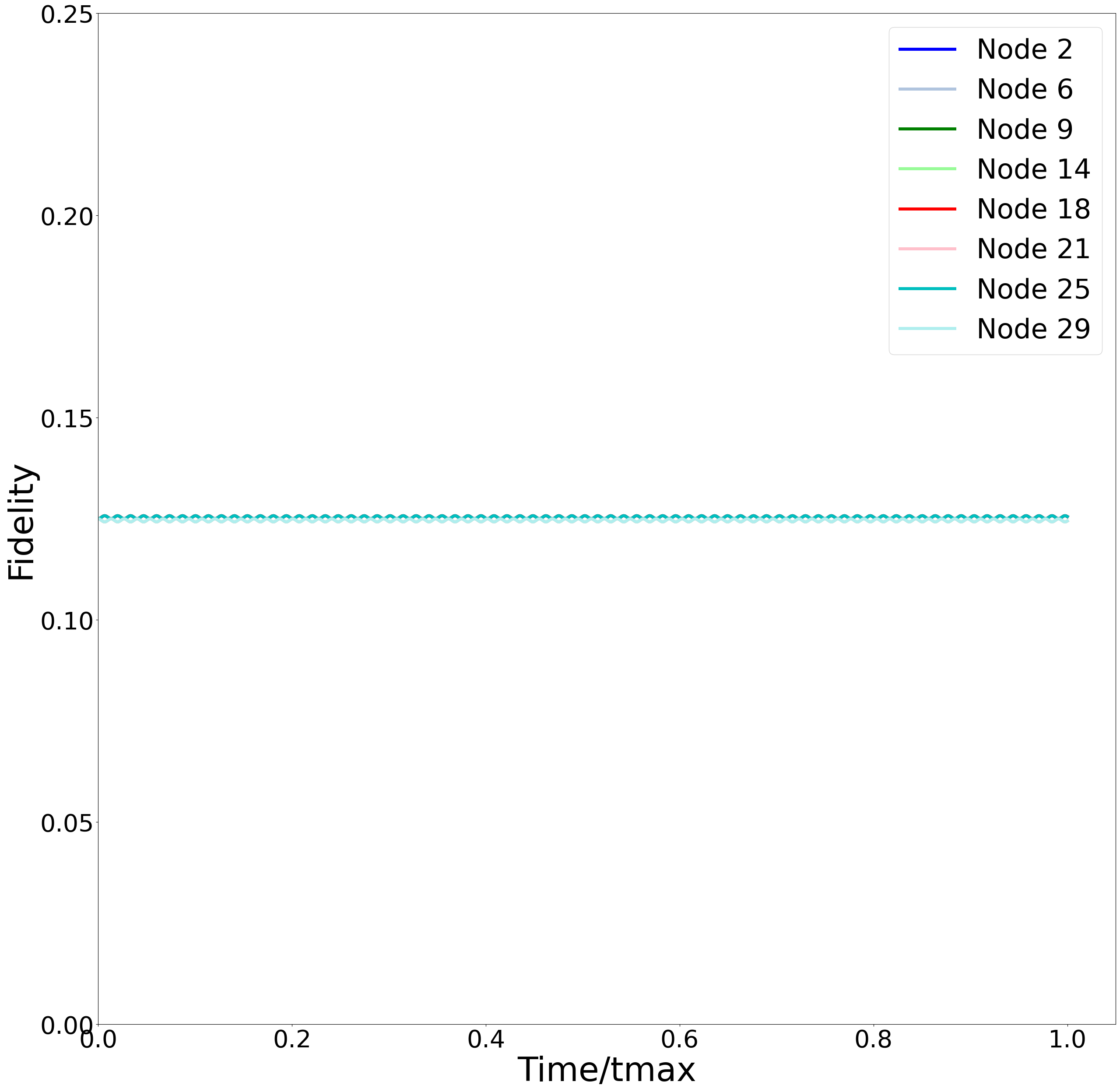}
            \caption{\small 
            coupling weights scaled like dipole-dipole interactions}
            \end{subfigure}
\caption{System dynamics of the MinMax network when initialised with the computational state.}
\label{fig:8N_MinMax_CompState}
\end{figure}

Initialising the system in this computational state gives rise to dynamics (Figure \ref{fig:8N_MinMax_CompState}) very similar to those seen in the Maximum Lengths network (Figure \ref{fig:8N_MaxLen_CompState}) in that the constant coupling scenario produces a static state in which all the lines are overlaid (due to an equal superposition being an eigenstate of the Hamiltonian) and there are periodic oscillations in the dipole coupling scenario corresponding to oscillating occupation probability between pairs of nodes coupled by a short (and therefore strong) connection. 
The oscillations in this case are much faster which is representative of the fact that the short connections in this case are much shorter in respect to the long connections between the unit cells. 

\subsection{Mid Lengths Network}
\subsubsection{Localised Initial State}
\begin{figure}[tp]
\centering
\begin{subfigure}[b]{\linewidth}
            \centering
            \includegraphics[width=0.91\textwidth]{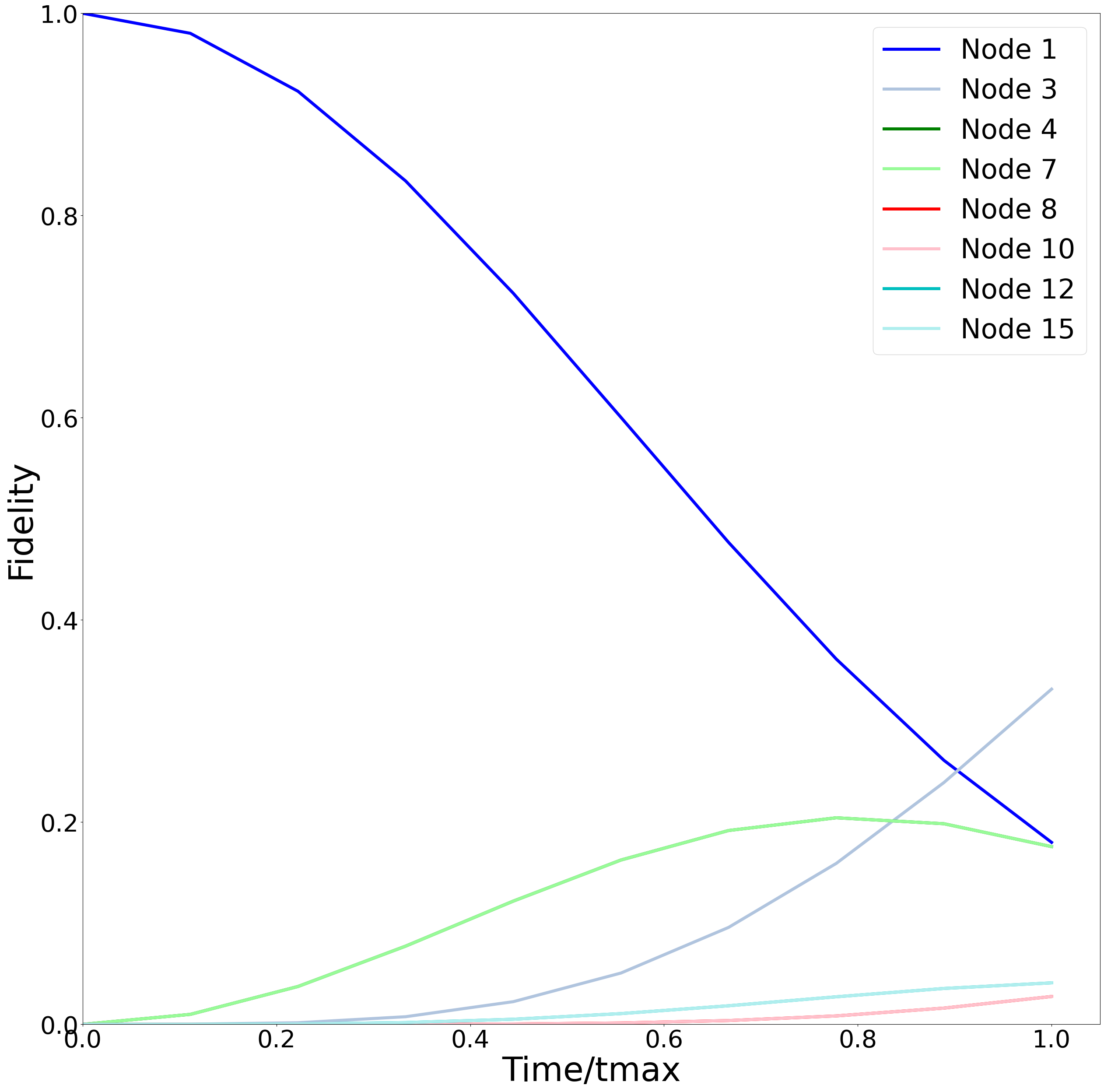}
            \caption{\small 
            coupling weights equal\\\vspace{1em}}
    \label{fig:IreneConstant}
            \end{subfigure}
        \\
        \begin{subfigure}[b]{\linewidth}  
            \centering 
            \includegraphics[width=0.91\textwidth]{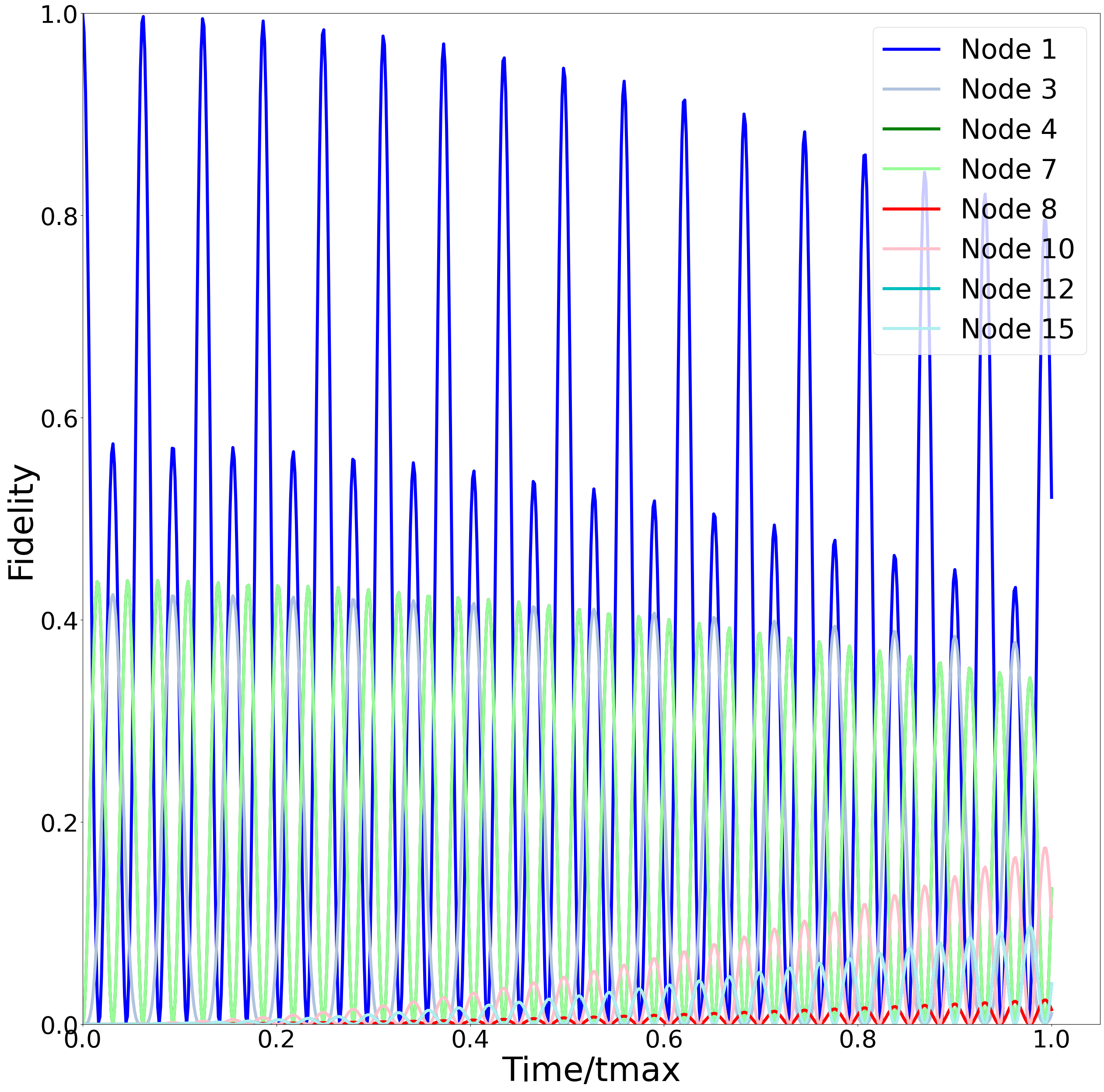}
            \caption{\small 
            coupling weights scaled like dipole-dipole interactions}
            \label{fig:IreneDipole}
            \end{subfigure}
\caption{System dynamics of the Mid Lengths network.}
\label{fig:8N_MidLen}
\end{figure}

The main difference in structure between this network (Figure \ref{fig:8N_Irene}) and the previous examples is the presence of subloops. The dynamics over the simulated time window when the initial state is a localised excitation on node 1 are shown in Figure \ref{fig:8N_MidLen}.
The constant coupling case still looks very similar to the previous graphs but in this case, the nodes that have overlaid lines are: nodes \#4 and \#7, nodes \#12 and \#15 and the pair \#8 and \#10. In previous examples, with no subloops, overlaid pairs represented nodes with the same minimum number of \enquote{hops} to the initial excitation node. In this case, this factor is combined with the number of routes of minimum length that the excitation can travel between the initial site and the node in question. This is why node \#3 behave unlike any other node in the system: like nodes \#12 and \#15, it is a minimum of two steps from node \#1 but in the case of node \#3, there are two different paths of two steps (either via node \#4 and via node \#7). 

The dipole case shows dynamics with a number of different frequencies. 
The site of initial excitation is connected to two other nodes by two connections of equal length which are also the smallest connections in the network.
This has the effect of a rapid dynamic of the fidelity splitting into equal halves across these connections and then rapidly transferring away to both the original node and node \#3. The peaks in node \#1 happen at the same time as those in \#3 but are larger in fidelity. This reflects the fact that the connection is shorter and therefore stronger. 
This is the dominant dynamic in the time scale considered here. However, across the time window, we can see increasing \enquote{leakage} to other nodes in the network via the longer connections. 

\begin{figure*}
\settoheight{\tempdima}{\includegraphics[width=.32\linewidth]{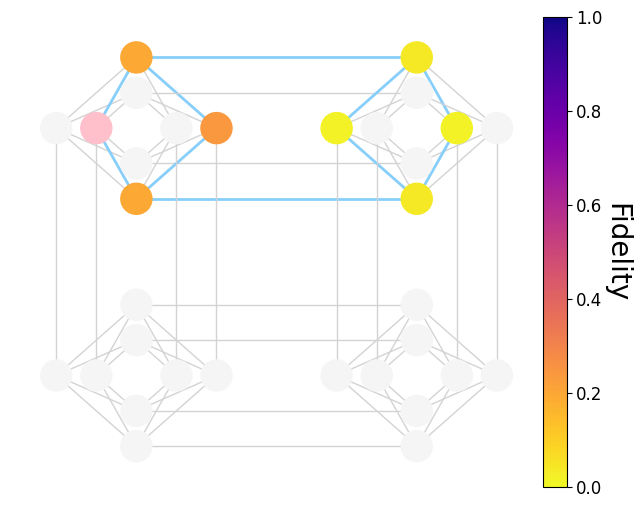}}%
\centering\begin{tabular}{@{}c@{ }c@{ }c@{}}
&\textbf{$t_{fP}$} & \textbf{$t_{mP}$}\\
\rowname{Constant}&
\includegraphics[width=.3\linewidth]{Results/Irene/const_first.png}&
\includegraphics[width=.3\linewidth]{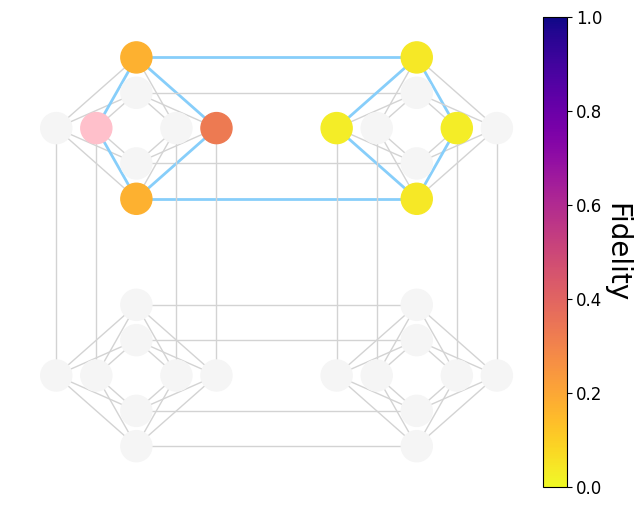}\\[-1ex]
&\textbf{(a)} & \textbf{(b)}\\
\rowname{Dipole}&
\includegraphics[width=.3\linewidth]{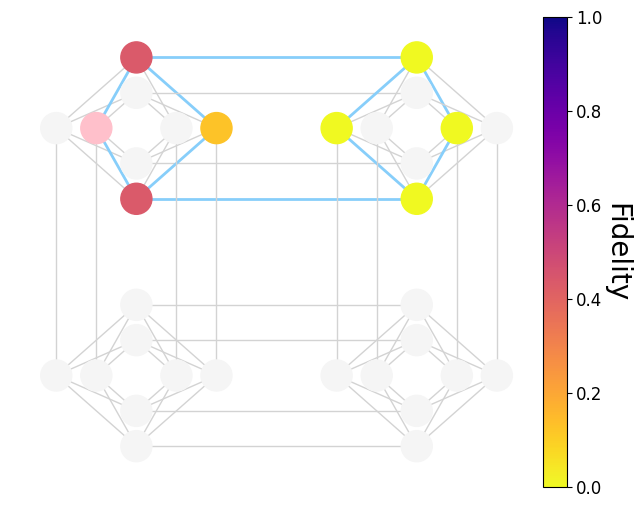}&
\includegraphics[width=.3\linewidth]{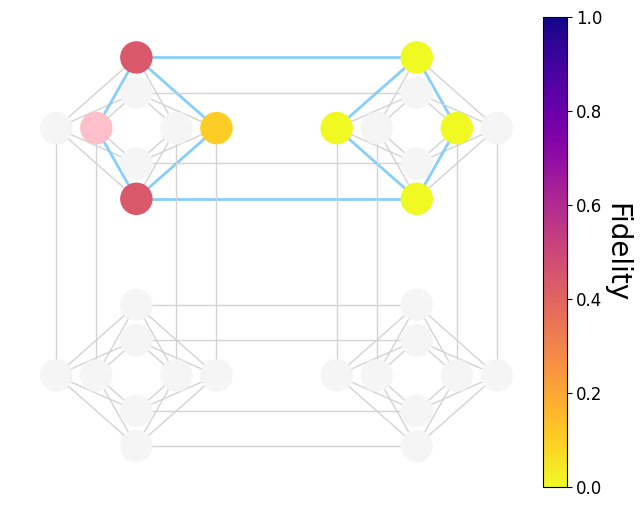}\\[-1ex]
&\textbf{(c)} & \textbf{(d)}\\
\end{tabular}
\caption{\small Node fidelities at two different times with two different couplings.}%
\label{fig:MidLenGraphs}
\end{figure*}

In all of the fidelity graphs, Figure \ref{fig:MidLenGraphs}, the excitation is isolated to the first unit cell. 
The constant coupling case produces a more even distribution of fidelity across this unit cell than the dipole case. 
In the constant case, node number 3 shows a fidelity approximately equal to or higher than the two node immediately connected to the original excitation location. 
This is likely because there is a a superposition of paths that lead to high occupation probability here. 
In the dipole case, the two immediately connected nodes show the highest fidelity in the network at both special times considered here (approximately 0.5). 
It would be beneficial to run this simulator over a longer time window to investigate whether the network would reach a steady state.

\begin{figure*}
\settoheight{\tempdima}{\includegraphics[width=.32\linewidth]{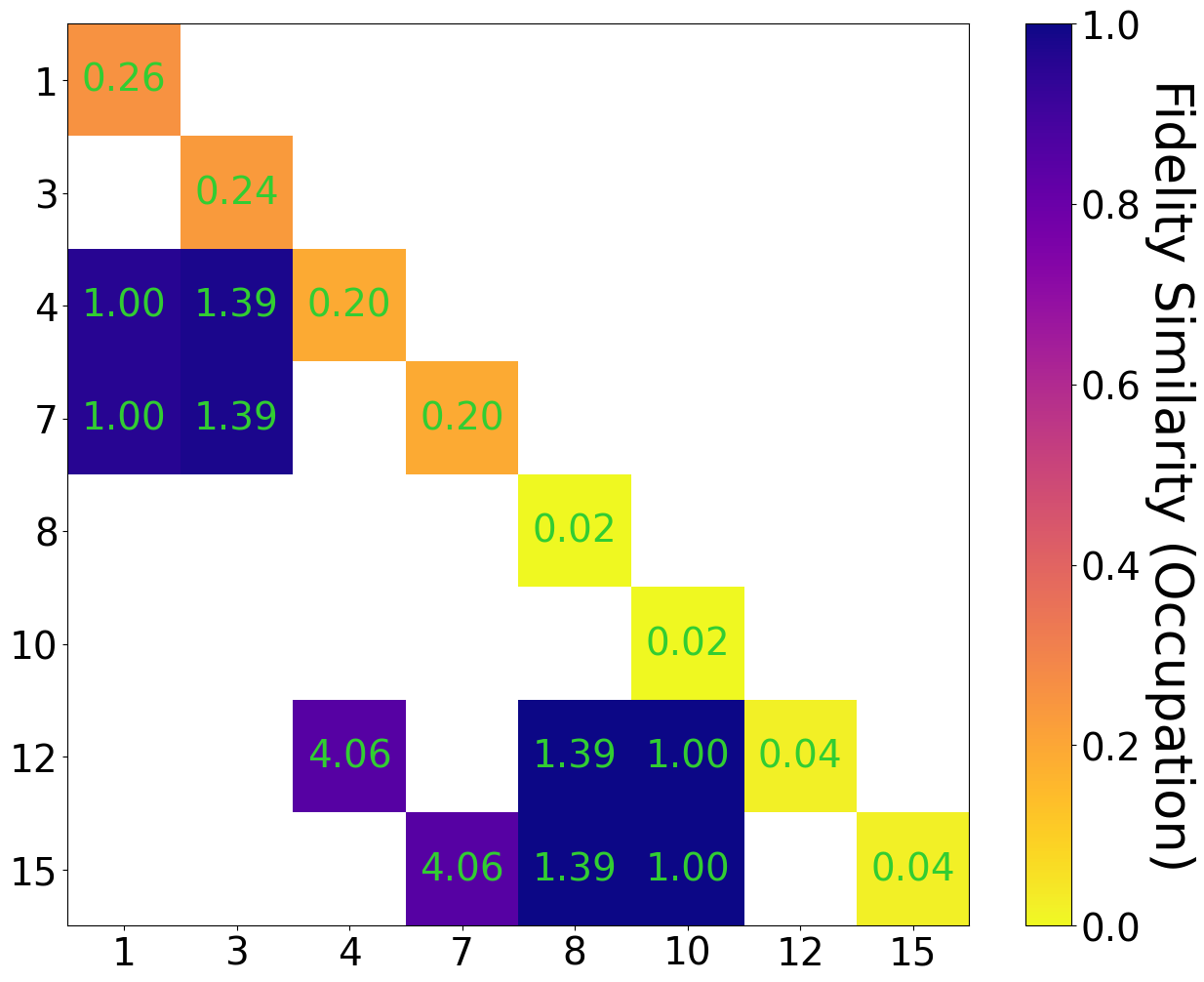}}%
\centering\begin{tabular}{@{}c@{ }c@{ }c@{}}
&\textbf{$t_{fP}$} & \textbf{$t_{mP}$}\\
\rowname{Constant}&
\includegraphics[width=.3\linewidth]{Results/Irene/const_first_grid.png}&
\includegraphics[width=.3\linewidth]{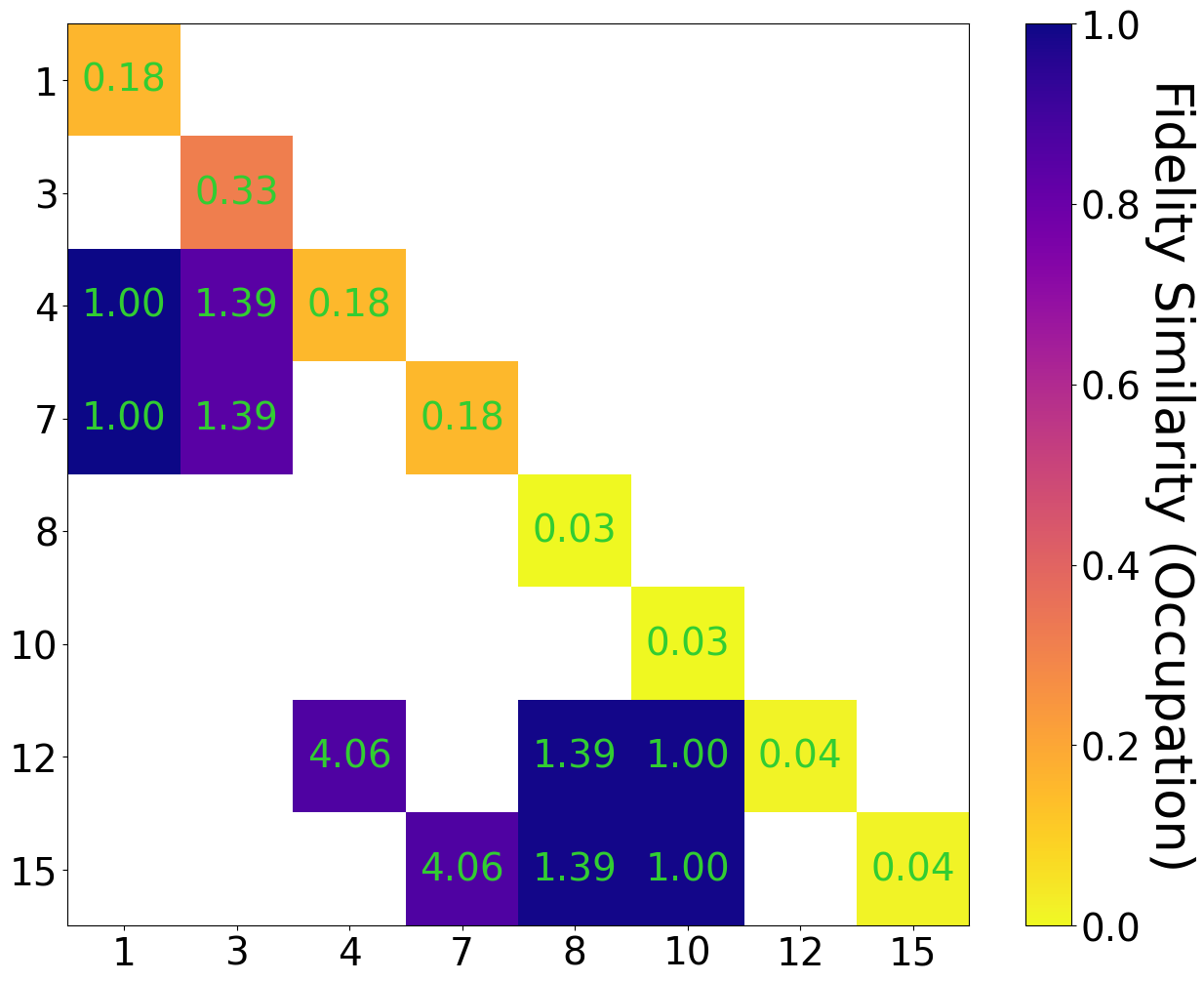}\\[-1ex]
&\textbf{(a)} & \textbf{(b)}\\
\rowname{Dipole}&
\includegraphics[width=.3\linewidth]{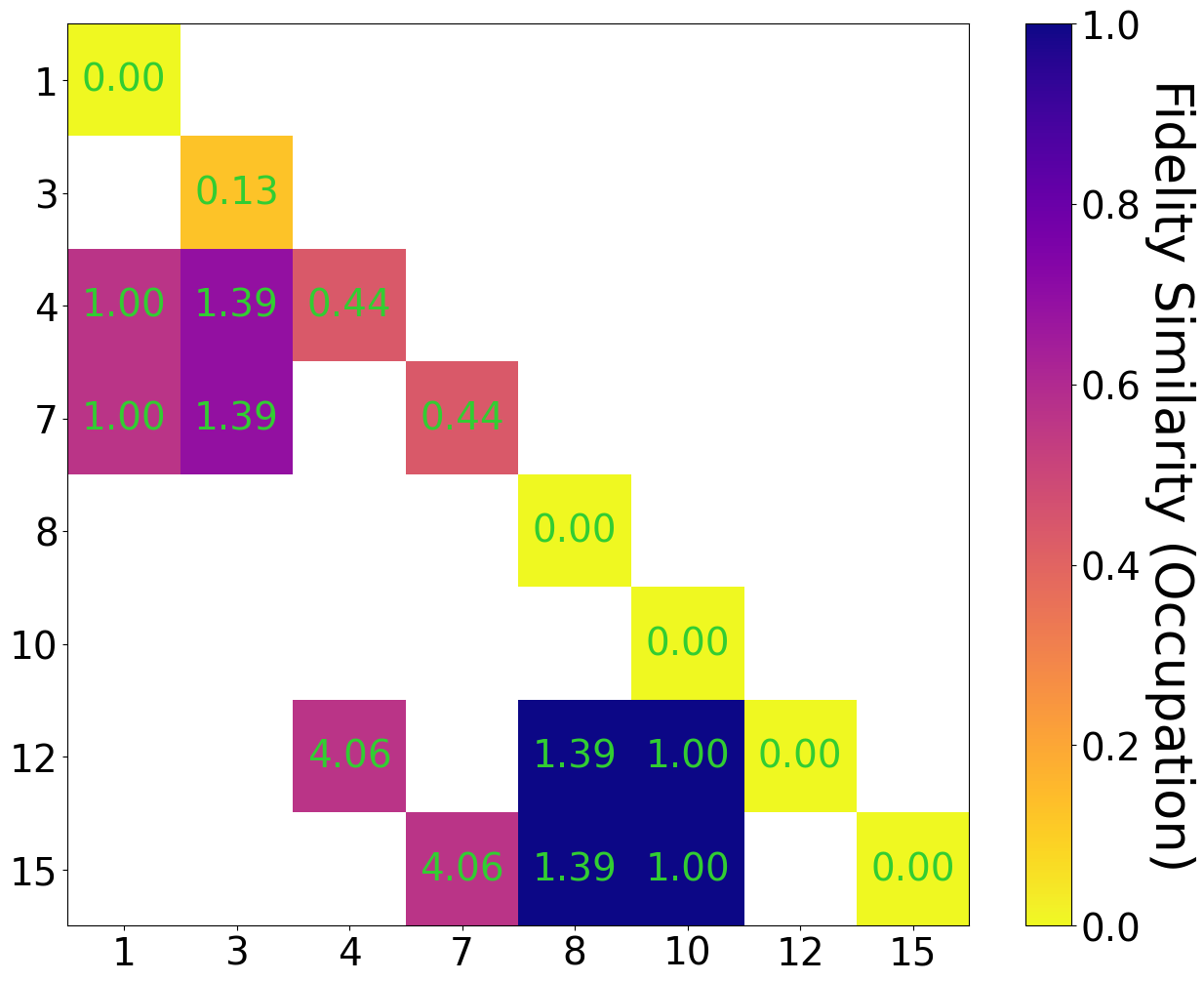}&
\includegraphics[width=.3\linewidth]{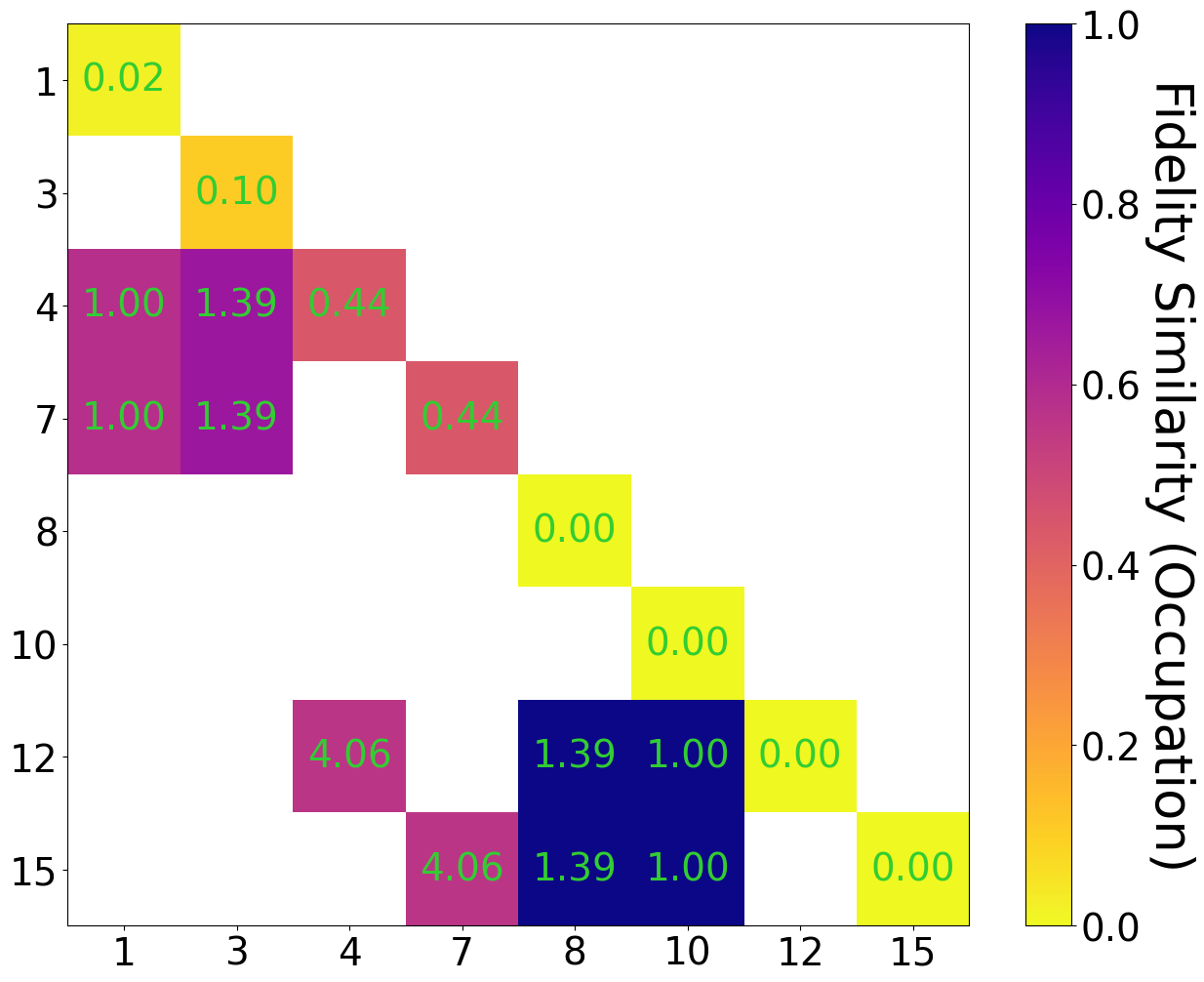}\\[-1ex]
&\textbf{(c)} & \textbf{(d)}\\
\end{tabular}
\caption{\small The similarity in the fidelity of connected nodes in the Mid Lengths Network. A square at ($i,j$) is labelled with the relative length of the connection between the nodes $i$ and $j$ and is coloured by the similarity as defined by Equation \ref{sim}. The diagonal is both labelled and colourised by that node's occupation fidelity at the time in question.}%
\label{fig:MidLenGrid}
\end{figure*}

The similarity graphs, Figure \ref{fig:MidLenGrid}, show high similarity within the individual unit cells. 
The left hand unit cell, has a fairly evenly spread occupation probability which in turn gives a high similarity value. 
In the righthand unit cell, all the nodes have fidelities very close to zero which also produces a high similarity value. 
The connections with a slightly lower similarity value are those which connect the two unit cells. 
The dipole cases also show high fidelity within the righthand unit cells due to all the nodes having fidelities close to zero. 
In the lefthand unit cell, the similarities are between 0.5 and 0.7 which is considerably higher than seen in the non-zero fidelity nodes in the other two networks considered in this paper. 
Therefore, we can suggest that the presence of subloops and many potential paths through the network, decrease the concentration of occupation fidelity at any one particular node. 
Increased occupation probability transfer would be useful when using these kind of networks for information transfer or computation. 

\subsubsection{Equal Superposition Initial State}
\begin{figure}[tp]
\centering
\begin{subfigure}[b]{\linewidth}
            \centering
            \includegraphics[width=0.91\textwidth]{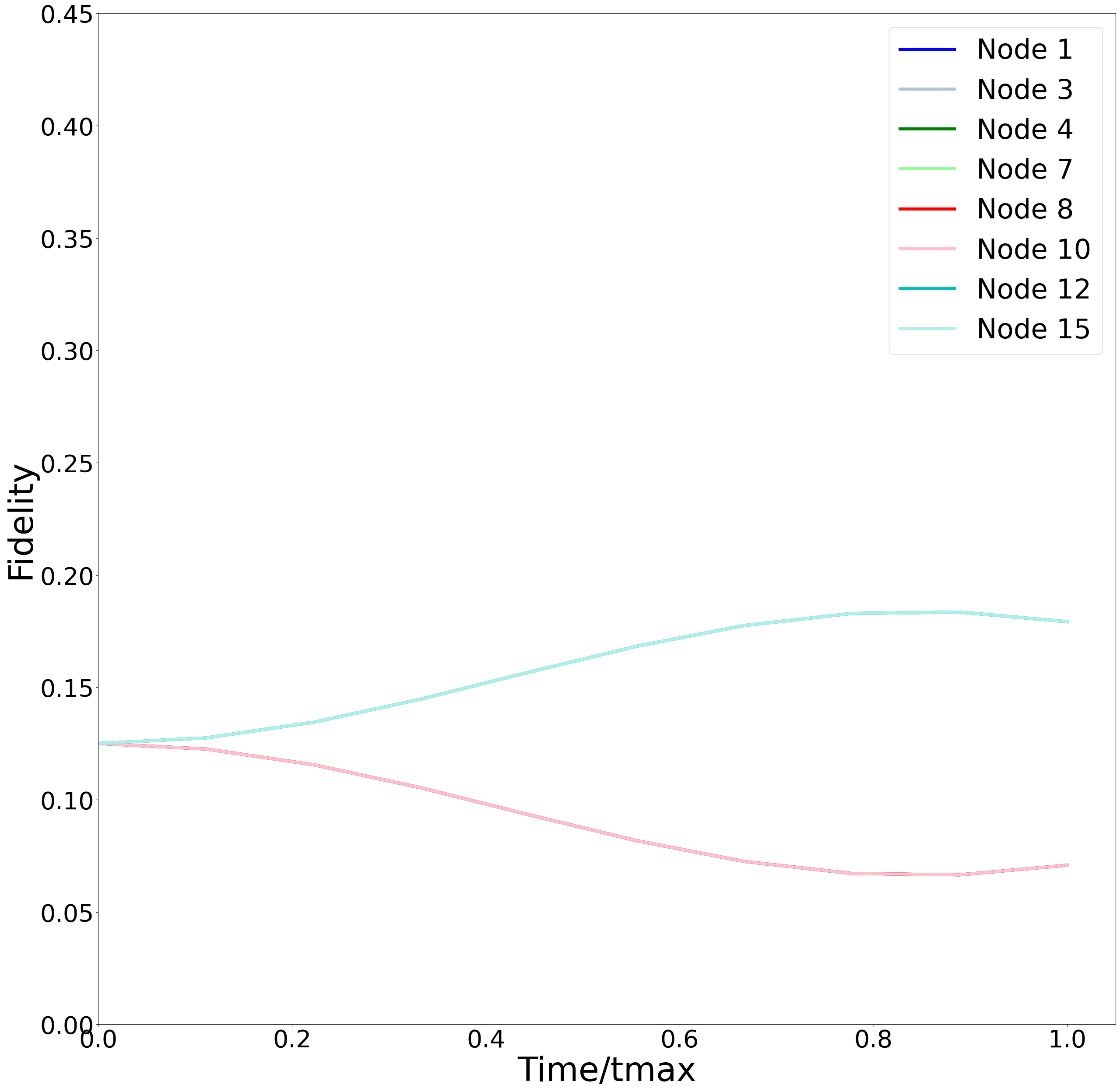}
            \caption{\small 
            coupling weights equal\\\vspace{1em}}
            \label{fig:Irene_Equ_Const}
            \end{subfigure}
        \\
        \begin{subfigure}[b]{\linewidth}  
            \centering 
            \includegraphics[width=0.91\textwidth]{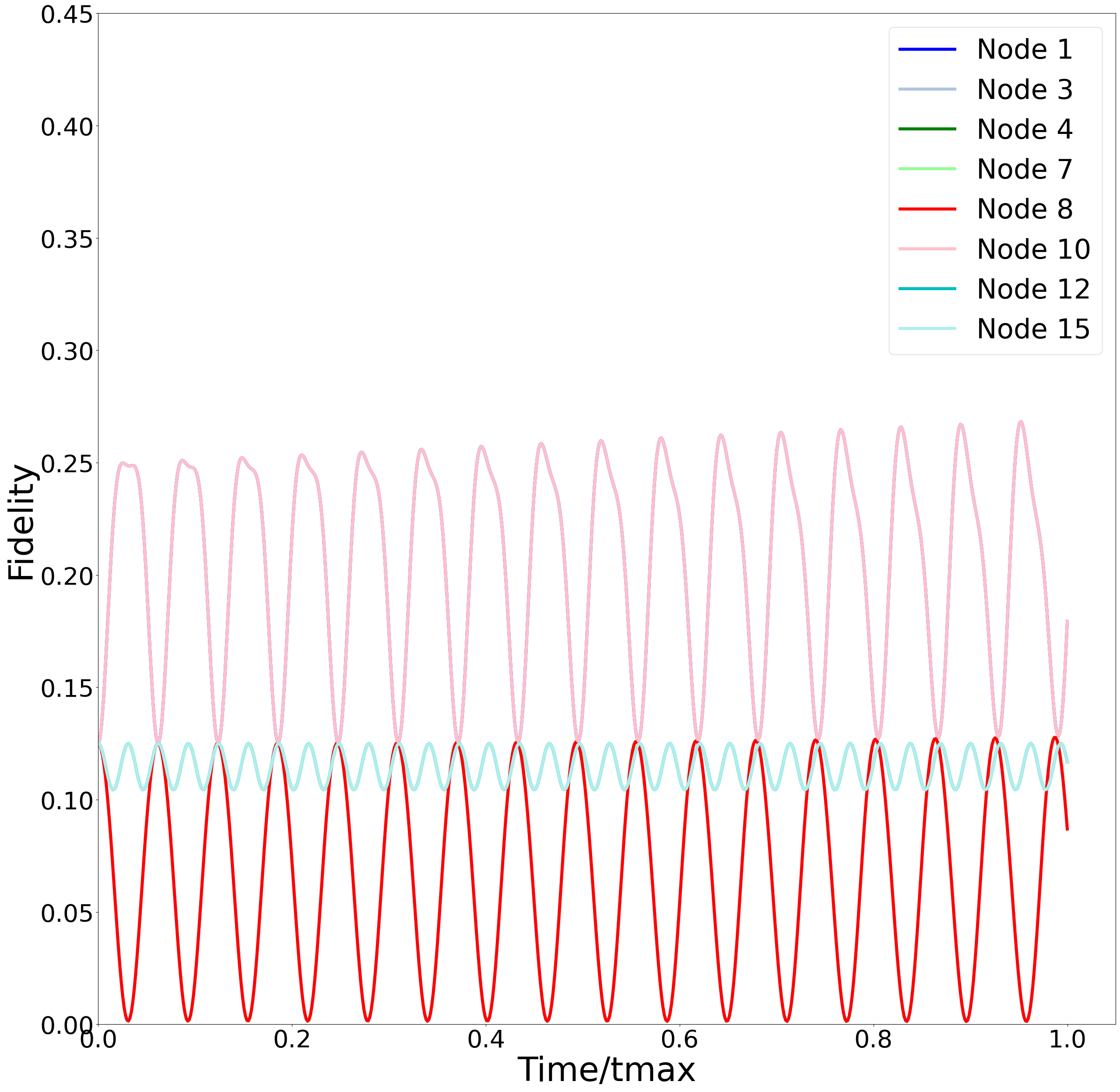}
            \caption{\small 
            coupling weights scaled like dipole-dipole interactions}
            \end{subfigure}
\caption{System dynamics of the Mid Lengths network when initialised with the computational state.}
\label{fig:8N_Irene_CompState}
\end{figure}

As mentioned, this network is distinct from the other two by the presence of subloops in the network. This is immediately seen by the fact that the constant coupling scenario doesn't induce a static state in this network when initialised with a equal superposition state (Figure \ref{fig:8N_Irene_CompState}). The equal superposition state is not an eigenstate of the Hamiltonian in this case and therefore dynamics are present. In this case the curves corresponding to nodes \#4, \#7, \#12 and \#15 are all overlaid onto the light cyan curve and the remaining nodes are all overlaid onto the pink curve. This does correspond to a difference between nodes that are connected to two others and those nodes connected to three other nodes. 
In the dipole-dipole coupling scenario there are further complicated dynamics. In this case, the nodes that are connected to three others (\#4, \#7, \#12 and \#15) show small-scale periodic oscillations (all overlaid onto the light cyan curve). The nodes connected to two others are now split into two groups, those with the shorter connections (\#1 and \#10) and those with the longer connections (\#3 and \#8) shown as the pink and red curves respectively). The pink curve shows consistently higher occupation fidelity than the red curve which corresponds to these shorter, and therefore stronger, connections). It is important to mention that the oscillation of the light cyan curve is of a different frequency to that of the pink and red curves. This causes a gradual sharpening of the peaks in the pink curve as it moves in and out of phase with the light cyan curve. 

\section{Conclusions and Future Work}\label{sec:Conc}

We have shown that there are strong positive spatial correlations in the qubits measured as part of the LANL study on single qubit fidelity beyond the horizontal/vertical delineation shown in the original paper \citep{Nelson2021-ca}. These correlations are only present in the connections between unit cells and not in those internal to unit cells. Similarly, we have shown the counterintuitive presence of negative spatial correlations between qubits internal to each unit cell. We hypothesise that this is due to the physical distances between the qubits affecting the connection strengths between them. 
More data, including both from the same device and from other D-Wave 2000Q chips, would be useful in determining whether these correlations seen here are a feature of the particular construction of this kind of chip or even if it is a repeatable phenomenon on exactly same chip. 

To provide evidence for this hypothesis, 
we created a simulated architecture of qubits (spins) with connection weights that depend on a variable scaling with the connection length, which we compared with the control case of constant coupling strengths across the network. We have considered the dynamics of a single excitation within the networks. Our results show that even when the couplings between the nodes are independent of length, the dynamics of the system are complex. We suggest that this is due to the connectivity of the network and the multiple paths an excitation could make to transfer from one node to another. Furthermore, when the connection strength is related to the physical distance between qubits, this has significant effects on the dynamics of the system beyond that due to the connectivity. We have introduced the concept of \enquote{similarity}, as a way of comparing the state of qubits directly connectd by an edge. This behaves in a complex (and sometimes counter-intuitive) way that is a combination of the effects due to connectivity and due to physical separation distance (when this is related to coupling strength). 

Different networks induce different frequencies of occupation probability transfer through the networks. When the coupling strengths are inversely proportional to the cube of the length of the connection (dipole-dipole interaction), having a larger difference between the shortest and longest connections in the network produces much faster dynamics along the short connections which in turn has the effect of preventing the information transfer throughout the network.
When the simulated network is comprised of multiple subloops, this changes the dynamics further and the excitation probability density tends to become more concentrated in more densely connected subloops as opposed to in single nodes as seen in the cases where the network is formed by a single loop. 

The differences between the effects of constant and dipole-dipole interaction, combined with the consequences of the connectivity of the network,
highlight the need to understand the effects of specific architectural features over those of the idealised model for progression quantum computation. 

When a system of spins with no subloops is connected by couplings of equal strength and is initialised in an equal superposition state, there are no dynamics through time. However, when the couplings between the spins have strengths that depend on the lengths of the connections, periodic oscillatory dynamics are induced in the system. The frequency of these oscillations is dependent on the spread of coupling strengths. 
When there are subloops in the network, even constant coupling strengths do not produce static fidelities on the nodes. In the third network tested here, the nodes don't all have the same number of neighbours, this is thought to be the source of the dynamics seen here.

The results shown here highlight the importance of understanding the underlying network architecture when designing quantum computing algorithms. If an algorithm requires 8 qubits, the choice of these 8 qubits could impact the performance of the algorithm. How the underlying network affects the algorithm will depend on the states used and the coupling strengths imposed on the network. 

An understanding of the effects seen here is important when benchmarking and comparing quantum devices and their performances on what might appear on the surface to be equivalent algorithms. 

When tackling large and dense problems on a quantum annealer, a process known as chaining is applied where two or more physical qubits are programmed to act as one logical qubit by fixing a large coupling strength between them. A simulator such as the one designed and used here could be used to test how successful chaining is especially in the scenario of unwanted interactions or environmental effects.


\section*{Declarations}

\subsubsection*{Ethical Approval} Not applicable, no human or animal research involved.

\subsubsection*{Competing Interests} The authors have no competing interests of a financial or personal nature, nor other interests that might be perceived to influence the results and/or discussion reported in this paper.

\subsection*{Authors' contributions} J.P. wrote the first draft of the manuscript and performed all simulation experiments. S.S. and I.d'A. reviewed the manuscript, and contributed to the design of the experiments and analysis of results. 

\subsubsection*{Funding} The authors wish to acknowledge Defence Science Technical Laboratory (Dstl) who are funding this research. 
Content includes material subject to © Crown copyright (2023), Dstl. This material is licensed under the terms of the Open Government Licence except where otherwise stated. To view this licence, visit http://www.nationalarchives.gov.uk/doc/open-government-licence/version/3 or write to the Information Policy Team, The National Archives, Kew, London TW9 4DU, or email: psi@nationalarchives.gov.uk

\subsubsection*{Data Availability}
The Los Alamos experimental data was kindly supplied by Carleton Coffrin. 
The code for the simulation experiments will be available on github.

\subsubsection*{Acknowledgments}

We thank Carleton Coffrin and his colleagues at the Los Alamos National Laboratory for sharing the data from their Single Qubit Fidelity Assessment.







\begin{appendices}






\end{appendices}


\bibliographystyle{apalike}
\bibliography{paperpile}

\begin{thebibliography}{}

\bibitem[Ahn et~al., 2002]{Ahn2002-pr}
Ahn, C., Doherty, A.~C., and Landahl, A.~J. (2002).
\newblock Continuous quantum error correction via quantum feedback control.
\newblock {\em Phys. Rev. A}, 65(4):042301.

\bibitem[Bandic et~al., 2022]{Bandic2022-it}
Bandic, M., Feld, S., and Almudever, C.~G. (2022).
\newblock Full-stack quantum computing systems in the {NISQ} era: algorithm-driven and hardware-aware compilation techniques.
\newblock In {\em 2022 Design, Automation \& Test in Europe Conference \& Exhibition ({DATE})}, pages 1--6. IEEE.

\bibitem[Bharti et~al., 2022]{Bharti2022-iu}
Bharti, K., Cervera-Lierta, A., Kyaw, T.~H., Haug, T., Alperin-Lea, S., Anand, A., Degroote, M., Heimonen, H., Kottmann, J.~S., Menke, T., Mok, W.-K., Sim, S., Kwek, L.-C., and Aspuru-Guzik, A. (2022).
\newblock Noisy intermediate-scale quantum algorithms.
\newblock {\em Rev. Mod. Phys.}, 94(1):015004.

\bibitem[{D-Wave Systems}, 2021]{D-Wave_Systems2021-bg}
{D-Wave Systems} (2021).
\newblock {{D-Wave} {NetworkX}}.

\bibitem[Geary, 1954]{Geary1954-cf}
Geary, R.~C. (1954).
\newblock The contiguity ratio and statistical mapping.
\newblock {\em The Incorporated Statistician}, 5(3):115--146.

\bibitem[Harris et~al., 2009]{Harris2009-qi}
Harris, R., Lanting, T., Berkley, A.~J., Johansson, J., Johnson, M.~W., Bunyk, P., Ladizinsky, E., Ladizinsky, N., Oh, T., and Han, S. (2009).
\newblock Compound {Josephson}-junction coupler for flux qubits with minimal crosstalk.
\newblock {\em Phys. Rev. B Condens. Matter}, 80(5):052506.

\bibitem[Mortimer et~al., 2021]{Mortimer2021-ex}
Mortimer, L., Estarellas, M.~P., Spiller, T.~P., and D'Amico, I. (2021).
\newblock Evolutionary computation for adaptive quantum device design.
\newblock {\em Advanced Quantum Technologies}, 4(8).

\bibitem[Nelson et~al., 2022]{Nelson2022-or}
Nelson, J., Vuffray, M., Lokhov, A.~Y., Albash, T., and Coffrin, C. (2022).
\newblock {High-Quality} thermal {Gibbs} sampling with quantum annealing hardware.
\newblock {\em Phys. Rev. Applied}, 17(4):044046.

\bibitem[Nelson et~al., 2021]{Nelson2021-ca}
Nelson, J., Vuffray, M., Lokhov, A.~Y., and Coffrin, C. (2021).
\newblock {Single-Qubit} fidelity assessment of quantum annealing hardware.
\newblock {\em IEEE Transactions on Quantum Engineering}, 2:1--10.

\bibitem[Noiri et~al., 2018]{Noiri2018-ym}
Noiri, A., Nakajima, T., Yoneda, J., Delbecq, M.~R., Stano, P., Otsuka, T., Takeda, K., Amaha, S., Allison, G., Kawasaki, K., Kojima, Y., Ludwig, A., Wieck, A.~D., Loss, D., and Tarucha, S. (2018).
\newblock A fast quantum interface between different spin qubit encodings.
\newblock {\em Nat. Commun.}, 9(1):5066.

\bibitem[Osada et~al., 2022]{Osada2022-os}
Osada, A., Taniguchi, K., Shigefuji, M., and Noguchi, A. (2022).
\newblock Feasibility study on ground-state cooling and single-phonon readout of trapped electrons using hybrid quantum systems.
\newblock {\em Phys. Rev. Research}, 4(3):033245.

\bibitem[Park et~al., 2023]{Park2023-zc}
Park, J., Stepney, S., and D'Amico, I. (2023).
\newblock Spatial correlations in the qubit properties of {D-Wave} {2000Q} measured and simulated qubit networks.
\newblock In {\em Unconventional Computation and Natural Computation}, pages 140--154. Springer Nature Switzerland.

\bibitem[Pudenz et~al., 2014]{Pudenz2014-oy}
Pudenz, K.~L., Albash, T., and Lidar, D.~A. (2014).
\newblock Error-corrected quantum annealing with hundreds of qubits.
\newblock {\em Nat. Commun.}, 5:3243.

\bibitem[Rey and Anselin, 2010]{Rey2010-uu}
Rey, S.~J. and Anselin, L. (2010).
\newblock {PySAL}: A {Python} library of spatial analytical methods.
\newblock In Fischer, M.~M. and Getis, A., editors, {\em Handbook of Applied Spatial Analysis: Software Tools, Methods and Applications}, pages 175--193. Springer.

\bibitem[Ronke et~al., 2011]{Ronke2011-rp}
Ronke, R., Spiller, T.~P., and D'Amico, I. (2011).
\newblock Effect of perturbations on information transfer in spin chains.
\newblock {\em Phys. Rev. A}, 83(1):012325.

\bibitem[Venegas-Andraca et~al., 2018]{Venegas-Andraca2018-yy}
Venegas-Andraca, S.~E., Cruz-Santos, W., McGeoch, C., and Lanzagorta, M. (2018).
\newblock A cross-disciplinary introduction to quantum annealing-based algorithms.
\newblock {\em Contemporary Physics}, 59(02):174--196.

\bibitem[Vuffray et~al., 2022]{Vuffray2022-nv}
Vuffray, M., Coffrin, C., Kharkov, Y.~A., and Lokhov, A.~Y. (2022).
\newblock Programmable quantum annealers as noisy gibbs samplers.
\newblock {\em PRX Quantum}, 3(2):020317.

\bibitem[Zhou and Lin, 2008]{Zhou2008-bj}
Zhou, X. and Lin, H. (2008).
\newblock Geary's {C}.
\newblock In Shekhar, S. and Xiong, H., editors, {\em Encyclopedia of {GIS}}, pages 329--330. Springer.

\end{thebibliography}

\end{document}